\documentclass[12pt,reqno,twoside,openright]{amsbook}

\usepackage[all]{xy}
\UseComputerModernTips
\CompileMatrices

\usepackage{amsmath}
\usepackage{amssymb}
\usepackage{latexsym}
\usepackage{amsthm}
\usepackage{eufrak}
\usepackage[dvips]{graphics}
\usepackage{setspace}
\usepackage{fontenc}
\usepackage{mathrsfs}
\usepackage{MnSymbol}
\usepackage{color}
\usepackage{manfnt}


\makeindex

\swapnumbers
\theoremstyle{plain}
\newtheorem{theorem}{Theorem}[chapter] 
\newtheorem{proposition}[theorem]{Proposition}

\theoremstyle{definition}

\DeclareMathOperator{\Sym}{Sym}

\DeclareMathOperator{\Aut}{Aut}

\DeclareMathOperator{\Alt}{Alt}

\DeclareMathOperator{\Spin}{Spin}

\DeclareMathOperator{\SAut}{SAut}

\DeclareMathOperator{\Cay}{Cay}

\DeclareMathOperator{\Cox}{Cox}

\DeclareMathOperator{\PSL(}{PSL(}
\DeclareMathOperator{\PGL(}{PGL(}

\DeclareMathOperator{\SL}{SL}

\newcommand{\sd}{\rtimes}

\DeclareMathOperator{\SO(}{SO(}

\DeclareMathOperator{\SU(}{SU(}
\DeclareMathOperator{\U(}{U(}
\DeclareMathOperator{\Bn}{Bn}

\DeclareMathOperator{\E_8}{E_8}
\newcommand{\Wr}{\mathop{\mathrm{Wr}}}
\DeclareMathOperator{\S(}{S(}

\begin{document}

\title{Multicoloured Random Graphs: The Random Dynamics Program}
\vspace{7cm}
\author{Sam Tarzi\\[3pt]
London, England}
\date{Draft, 15 June 2014}
\maketitle

\pagenumbering{roman}


\vspace{3cm}


\begin{titlepage}
\rule{0pt}{3cm}
\renewcommand{\baselinestretch}{2}

\end{titlepage}

\chapter*{Copyright Notice}

\bigskip

\centerline{\copyright\  2014 Sam Tarzi}
\centerline{All Rights Reserved}

\vspace{1cm}

\begin{flushleft}
Sam Tarzi is identified as the author of this work in accordance with the Copyright, Designs and Patents Act, 1988.
\end{flushleft}

\vspace{1cm}

\begin{flushleft}
No part of this publication may be reproduced or stored in a retrieval system, nor transmitted in any form or by any means - digital scanning, mechanical, photo-copying, recording, or by any other media - without prior permission of the copyright owner.
\end{flushleft}

\vspace{1cm}

\begin{flushleft}
Every effort has been made to trace copyright ownership and permissions are confirmed in the text.  Any omission brought to the notice of the publishers will be acknowledged in future impressions.
\end{flushleft}

\clearpage



\chapter*{Abstract}

\centerline{Multicoloured Random Graphs: The Random Dynamics Program}

\bigskip

The Random Dynamics
\index{Random Dynamics}%
 program of Holger Nielsen
\index{Nielsen, H. B.}%
 and co-workers is a proposal to explain the origin of all symmetries, including Lorentz and gauge invariance
\index{Lorentz invariance}%
\index{gauge ! invariance}%
 without appeal to any fundamental invariance of the laws of nature, and to derive the known physical laws in such a way as to be almost unavoidable.

The philosophy of Random Dynamics
\index{Random Dynamics}%
 is that the most useful assumption that we can make about the fundamental laws is that they are random and then to see how known physics like mechanics and relativity follow from them.  It is believed that almost any model of these regularities would appear in some limit for example as energies become small.   Almost all theories or models at the fundamental level could then explain known physics and specific models would arise from chaotic fundamental physics which has as few starting assumptions as possible.

This brief book continues the previous two in the series, the first being a derivation of pure mathematical results and the second containing a sketch of possible physical applications of some of the results in the first volume.  In this third volume we focus on the ideas behind the Froggatt-Nielsen project and suggest how using the formalism and properties of random graphs can be useful in developing the theory, and point towards directions in which it can be more fully extended in future work.

In the previous volume we sketched our proposal for the origin of the Standard Model
\index{Standard Model (SM)}%
 family structure. Here we look in greater detail at some of the other proposals of the Random Dynamics
\index{Random Dynamics}%
 program and outline possible mathematics and in particular properties of multicoloured random graphs
\index{graph ! random ! $m$-coloured}%
 as possible starting points in a research project that applies the mathematics to the physics by judicious interpretation of the various elements of the random models as physical concepts.  We would hope that the ideas presented here could be implemented into a detailed formal theory, as part of an exploration of the graph-theoretic and related group-theoretic content of Random Dynamics.
\index{Random Dynamics}%
  In this prelude to an investigation of the mathematics that may be applied to the Random Dynamics program we retain the nomenclature used by Nielsen and co-workers.

\clearpage

\chapter*{Dedication}


\vspace{4cm}

\centerline{\textbf{\Large In loving memory of my late father Anwar}} 

\bigskip

\centerline{\textbf{\Large my mother Adiba}}

\bigskip

\centerline{\textbf{\Large Jacob Avner and Daniel}}

\clearpage

\tableofcontents

\bigskip

\clearpage

\pagenumbering{arabic}

\chapter{Introduction and Definitions}

The purpose of this book is to publicize both the Random Dynamics
\index{Random Dynamics}%
 program as well as a possible approach to it using the theory of multicoloured random graphs, by giving a flavour of the properties of these graphs together with a hint of the associated mathematics.
\index{graph ! random ! $m$-coloured}%

The Random Dynamics project is an approach to physics that differs from the usual bottom-up approach of building on known theory by unifying different phenomena into one theory.  It is in the spirit of Smolin's
\index{Smolin, L.}%
 clearly-argued~\cite{smolintr} need for radical rather than incremental approaches to furthering our understanding of how things work.

Its premise is that all symmetries, including the most fundamental ones such as Lorentz
\index{Lorentz invariance}%
 and gauge invariance,
\index{gauge ! invariance}%
 can be derived without having to be assumed, even if the most fundamental laws at very high energies do not possess them. 

Many symmetries such as charge conjugation, parity and strangeness are considered to arise from the \emph{gauge field theories}
\index{gauge ! theory}%
 of electromagnetic and strong interactions, whilst others arise from taking the limits in physical models such as for small mass-energies.  The \emph{gauge group}
\index{group ! gauge}%
 is a group of gauge transformations
\index{gauge ! transformation}%
 between the possible gauges.  
\index{gauge}%

\emph{Gauge symmetry}
\index{gauge ! symmetry}%
 itself is perhaps not a true symmetry but a redundancy in our descriptions.  However as well as its use in build theories it has been a mechanism in explaining fundamental issues such as why the weak charge depends on handedness and why there is parity violation in the Standard Model
 (SM)?
\index{Standard Model (SM)}%
 It has been argued~\cite{frog} that gauge invariance protects quarks and leptons from gaining a fundamental mass that is large compared to the electroweak scale, and so their gauge charges must depend on handedness; furthermore parity conservation in the electromagnetic and strong interactions is derived.

The Standard Model
\index{Standard Model (SM)}%
 comprises three leptons $(e, \mu, \tau)$, three neutrinos $(\nu_e, \nu_{\mu}, \nu_{\tau})$, three up-type quarks $(u, c, t)$, three down-type quarks $(d, s, b)$.  These undergo gauge field theory
\index{gauge ! theory}%
 interactions mediated by the force carrier bosons, the photon, gluons and $W$ and $Z$ bosons.  Particles that have mass gain it through a condensing Higgs boson.

Approximate SM symmetries such as chiral symmetry and Gell-Mann and Ne'eman's
\index{Gell-Mann, M.}%
\index{Ne'eman, Y.}%
 $\SU(3)$ flavour symmetry~\cite{gellnee} can be derived in the limit as quark masses become ever smaller~\cite{weinb}.

The handedness of a particle gives the direction of its spin vector along the direction of motion, and the SM prediction that neutrinos always have left-handed spin means that its spin vector always points in the opposite direction to its momentum vector.  If there were only left-handed neutrinos, as required by the SM, then they would be massless for according to special relativity a neutrino with mass cannot travel at the speed of light.  Then a faster moving observer could therefore overtake the spinning massive neutrino and would see it moving in the opposite direction and thus it would thus appear right-handed.  So the non-zero neutrino mass and mixing that has been observed is evidence for new physics beyond the SM.  Although the tri-bimaximal mixing pattern may now have been excluded, the discrete family symmetry approach is still favoured after experimental measurements, it could still be a possible good first approximation to the observed data.  The alternating group of even permutations of 4 objects $\Alt(4)$
\index{group ! Alt(4)@$\Alt(4)$}%
is the simplest symmetry group to leading order that leads to tri-bimaximal mixing.  Altarelli, Feruglio and Merlo's
\index{Altarelli, G.}%
\index{Feruglio, F.}%
\index{Merlo, L.}%
speculation on the origin of $\Alt(4)$
\index{group ! Alt(4)@$\Alt(4)$}%
  as a subgroup of the modular group
\index{group ! modular}%
 $\PSL(2, \mathbb{Z})$
\index{group ! PSL(2, \mathbb{Z})@$\PSL(2, \mathbb{Z})$}%
 is in line with our suggestion that discrete sugroups in particle physics may arise as homomorphic images of this group.  King and Luhn
\index{King, S. F.}%
\index{Luhn, C.}%
 have written a recent review article on discrete family symmetry approaches to modeling neutrino mass and mixings~\cite{king}.

On his website, Holger Nilesen,
\index{Nielsen, H. B.}%
 the originator of the Random Dynamics
\index{Random Dynamics}%
 philosophy states:

``In the Random Dynamics scheme the predicament of the poor physicist is taken very seriously, as it is postulated that we simply cannot deduce what the physics is like at the highest energy levels. Therefore, in order to avoid carrying along the usual tacit assumptions in fabricating a theory of the fundamental physics, we consider a most general, ``generic''
\index{structure ! generic}%
 mathematical structure which we equip with some general properties and notions. The ambition is to create a (as explicit as possible) list of assumptions, in order to pin down what conceptual luggage we cannot do without, if we have the ambition of formulating a theory for the fundamental physics, which at our low energy level appears as the physics we observe, as described by the Standard Model.'' (http://www.nbi.dk/~kleppe/random/////qa/poor.html)
\index{Standard Model (SM)}%

One motivation for the top-down Random Dynamics
\index{Random Dynamics}%
 program are the laws of thermodynamics which were at first built upon certain principles but are now understood to be a reflection of the behaviour of large numbers of randomly-moving particles.

Symmetry has been identified as a key theme in understanding the principles underlying natural phenomena in theoretical physics.  It represents the idea that observed physics is invariant under some symmetry transformation. 

The Grand Unification program in particular postulates that a larger gauge group
\index{group ! gauge}%
 beyond that relevant to the Standard Model (SM)
\index{Standard Model (SM)}%
 corresponds to a particular force and that at lower energies this group breaks down to the SM group corresponding to the forces already discovered.  Random Dynamics
\index{Random Dynamics}%
 assumes that any extension of the SM cannot be inferred from SM itself.

The Random Dynamics
\index{Random Dynamics}%
 project requires starting from general assumptions about an abstract relational mathematical construct and using very generic arguments because the idea is that the details of the very complicated laws are almost unimportant, only that we study them in some limits such as that of low energy.  We take seriously their proposal to begin with a random mathematical structure and put forward evidence as to why the family of multicoloured random graphs
\index{graph ! random ! $m$-coloured}%
 provide a concrete possibility.

\bigskip

A \emph{simple, undirected graph}
\index{graph ! simple}%
\index{graph ! undirected}%
 is an ordered pair $\Gamma = (V, E)$ consisting of a set $V$ of vertices, and a set $E$ of edges or lines which are $2$-element subsets of $V$ so that an edge is a relation on pairs of vertices.  A \emph{complete graph}
\index{graph ! complete}%
is one which has an edge on every pair of vertices, and its complement the \emph{null graph}
\index{graph ! null}%
is one which has only non-edges on every pair of vertices.  An \emph{induced subgraph} is one obtained by deleting vertices and the incident edges.  
 A \emph{spanning graph}
\index{graph ! spanning}%
 is a subgraph of a graph that is a spanning tree (having no cycles) containing all the vertices of the over-graph.  A bijection $\alpha: V(\Gamma) \to V(\Gamma')$ between the vertex sets of two graphs $\Gamma$ and $\Gamma'$ is an \emph{isomorphism} if $\{u, v\} \in E(\Gamma) \Leftrightarrow \{\alpha(u), \alpha(v)\} \in E(\Gamma')$.  If $\Gamma = \Gamma'$, then the isomorphism is an \emph{automorphism}.
\index{graph ! automorphism}%
 
An \emph{automorphism}
\index{graph ! automorphism}%
of a graph is a permutation of the vertex set preserving adjacency.  

Choose a \emph{natural number} $m \in \mathbb{N}$, $m \ge 2$, and a \emph{countable} (that is finite or countably infinite) set $V$ of vertices and colour the edges of the complete graph
\index{graph ! complete}%
 on $V$ from $m$ colours $c_1, \ldots, c_m$, selecting colours independently for each edge at random.  By a random colouring of a graph we simply mean an arbitrary colouring.

Denote $m$-coloured random graphs by $\mathfrak{R}_{m, \omega}$,
\index{graph ! random ! $m$-coloured}%
 where $m \in \mathbb{N}$, $m \ge 2$, and $\omega$ denotes the countably infinite vertex set.
\index{graph ! random ! $m$-coloured}%
 As we will only consider infinite graphs we can omit $\omega$.  They are defined by an infinite iteration of the following \emph{one-point extension property} for $m$ edge-coloured graphs:

$(*_m)$ 	Given finite disjoint sets $U_1, \ldots, U_m$ of vertices there is a vertex $z$ such that all edges from $z$ to $U_i$ have colour $c_i$ $(i=1, \ldots, m)$.

For $m = 2$ the colours are conventionally characterized as edges and non-edges.  The random two-coloured graph $\mathfrak{R}$ has a countably infinite number of vertices.   Its most common variant is as a simple graph, thus having only two possible adjacencies with no loops on single vertices or multiple edges between pairs of vertices being permitted.  It is a remarkable fact~\cite{tarzi1} that a countable two-coloured random graph
\index{graph ! random ($\mathfrak{R}$)}%
  is isomorphic to $\mathfrak{R}$ with probability $1$.   

We denote the countable random graph with three adjacency types, which we call the \emph{triality graph},
\index{graph ! triality ($\mathfrak{R^{t}}$)}%
by $\mathfrak{R^{t}}$.  We can get from $\mathfrak{R}_{m}$ to $\mathfrak{R}$ by going colourblind in pairs of colours consecutively until there are only two colours left.

\bigskip

There are three equivalent ways of characterizing $\mathfrak{R}$ up to isomorphism as a countable graph, these being the so-called one-point extension property, the injection property
\index{injection property}%
 or a combination of universality and homogeneity~\cite{tarzi1}.

A \emph{universal} structure
 \index{structure ! universal}%
 is one that contains all finite substructures, and a homogeneous structure
 \index{structure ! homogeneous}%
is one for which any isomorphism between finite substructures extends to an automorphism of the entire structure.  To say that a structure is homogeneous
\index{structure ! homogeneous}%
 is equivalent to asserting that it has the maximum amount of symmetry; or that many parts of the structure look alike; or that the structure looks the same when viewed from many positions within the structure.

We have mentioned the uniqueness of the countable random graph;
\index{graph ! random ($\mathfrak{R}$)}%
 it also has a huge amount of symmetry.  Most graphs, certainly finite ones, are \emph{asymmetric},
\index{graph ! asymmetric}%
that is they possess a trivial automorphism group
\index{group ! automorphism}%
 consisting solely of the identity element.   By contrast the automorphism group of $\mathfrak{R}$ and its multicoloured
\index{graph ! random ! $m$-coloured}%
 versions are gigantic, being uncountably infinite in size, thereby indicating a colossal degree of symmetry possessed by these graphs.

To see how remarkable and unexpected this is, contrast the situation with the finite case.  Construct a finite random graph on $n$ vertices by ordering the pairs of vertices in a countable sequence, and tossing a fair coin a total of $\frac{n(n-1)}{2}$ times where heads means join the two vertices by an edge, and tails means do nothing.  Every $n$-vertex graph (up to isomorphism) occurs with non-zero probability, and the probability of a particular graph arising is inversely proportional to its symmetry as measured by the order of its automorphism group.
\index{group ! automorphism}%
 The asymmetric graphs
\index{graph ! asymmetric}%
 are overwhelmingly more numerous than the symmetric ones.   The probability measure associated with a countable sequence of coin tosses is discussed in Category and Measure Appendix in~\cite{tarzi1}.  To get a sense for the meaning of homogeneity beyond its definition consider that the pentagon or 5-cycle graph is homogeneous, but that the hexagon or 6-cycle graph is not because a pair of points two steps apart and a pair three steps apart are isomorphic as induced subgraphs but are not equivalent under automorphisms of the whole graph.
  
\smallskip

The existence and uniqueness of homogeneous universal structures follow immediately from Fra\"{\i}ss\'e's Theorem~\cite{f} on relational structures.
\index{Fra\"{\i}ss\'e's Theory}%
\index{structure ! relational}%
The basic concept in Fra\"{\i}ss\'e's work 
\index{Fra\"{\i}ss\'e, R.}%
is the \emph{age}
\index{structure ! age}%
Age$(\mathcal{M})$ of a relational structure $\mathcal{M}$, which is the class of all finite structures (over the same logical language) which are embeddable in $\mathcal{M}$ as induced substructures;  see the Appendix on Fra\"{\i}ss\'e's Theory
\index{Fra\"{\i}ss\'e's Theory}%
 of Relational Structures in~\cite{tarzi1}. As we stated, homogenous structures are those with the maximum amount of symmetry and one of the most useful methods for constructing objects with a large amount of symmetry is based on this theorem.  In the two-colour case, thinking of a
two-coloured complete graph as a simple graph, the appropriate structure is
\emph{Rado's graph},
\index{graph ! Rado's}%
or
the Erd\H{o}s-R\'enyi
\index{R\'enyi, A.}%
\index{Erd\H{o}s, P.}%
 \emph{random graph}
\index{graph ! random ($\mathfrak{R}$)}%
~\cite{er} $\mathfrak{R}$; it
was Richard Rado
\index{Rado, R.}%
~\cite{rado} who gave the first construction of
$\mathfrak{R}$. 

Relational structures
\index{structure ! relational}%
 with large automorphism groups
\index{group ! automorphism}%
 have a high degree of
symmetry; this is also implied by an orbit space with a small number of orbits of the
acting automorphism group.  A countable homogeneous relational
structure
\index{structure ! homogeneous}%
 of a given type is determined up to isomorphism by the
isomorphism classes of its finite substructures.  An important outcome of Fra\"{\i}ss\'e's Theorem
\index{Fra\"{\i}ss\'e's Theory}%
 is that determining the numbers of orbits on $n$-sets or $n$-tuples of \emph{oligomorphic} permutation groups
\index{group ! permutation ! oligomorphic}%
(those having finitely many orbits in their induced action on $n$-tuples for all $n$) is equivalent to enumerating the unlabelled or labelled objects in certain classes of finite structures characterized largely by the \emph{amalgamation property}, which is a way of building larger structures from smaller ones;  see the Appendix on Fra\"{\i}ss\'e's Theory
\index{Fra\"{\i}ss\'e's Theory}%
 of Relational Structures in~\cite{tarzi1}.  Not all graph classes amalgamate; for example, $n$-colourable graphs (those having chromatic number at most $n$) do not have the amalgamation property. 

The word \emph{oligomorphic}
\index{group ! permutation ! oligomorphic}%
is intended to capture the notion of `few shapes', where the group orbits contain a finite number of structures of any given finite size (that is, few) and each orbit contains an isomorphism class of structures (that is, shapes).

\bigskip

It is a fair assumption that the fundamental principles underlying physics are simple conceptions, whilst the phenomena themselves can be very complicated.  In Random Dynamics
\index{Random Dynamics}%
 it is postulated that this complexity is best described through randomness.

The ambitious goal of Random Dynamics
\index{Random Dynamics}%
 is to \emph{derive} all the known physical laws as an almost unavoidable consequence of a random system which it is assumed would be a very general, random mathematical structure containing non-identical elements and set-theoretical notions such as subsets.  It is proposed that the physics of our experience is derived from this basic mathematical structure, but as we do not have access to fundamental scales such as the Plack scale the requisite mathematical object(s) should have a generic structure,
\index{structure ! generic}%
 without a priori notions of distance, geometry, differentiability etc.  The `random' aspect of the dynamics represents our lack of knowledge.

The infinite random graphs
\index{graph ! random ($\mathfrak{R}$)}%
 are simply defined and their open definition leaves much room for enhancing the graphs with a myriad of properties such as different topologies, required to model a particular physical fact.  The surprising uniqueness (up to isomorphism) of the countably infinite random graph for a given number of edge-colours is of course reminiscent of this philosophy; it means that given \emph{any} finite graph, as more and more vertices are attached the eventual infinite graph is almost always isomorphic to $\mathfrak{R}$.  
\index{graph ! random ($\mathfrak{R}$)}%

If even space and time are to be derived then paths in our connected random graphs
\index{graph ! random ($\mathfrak{R}$)}%
 may approximate the discrete time-slicing of points in the path integral
\index{path integral}%
 formulation of quantum mechanics (QM) and thereby give the principle of locality in QM.  It may be that there are graph-theoretic criteria on the allowed paths in the path integral formulation or the decoherent histories interpretation of QM that narrow the possibilities for the sets of projective decompositions to which the histories correspond, and thereby help in the formulation and solution of problems; see the section on quantum locality in~\cite{tarzi2}, where criticisms by Bassi,
\index{Bassi, A.}%
 Kent
\index{Kent, A.}%
 and Weinberg
\index{Weinberg, S.}%
 are mentioned together with references.

It would be advantageous if the graphs are supplemented with as little extra information as possible, and certainly properties such as topologies can be defined from the graph structure itself.  Some of the mathematics, such as probability measures on the graphs, is already worked-out in the literature (see the Appendix on Category and Measure in~\cite{tarzi1}) and some will can be used to describe the observed world.

The requirement that every useful physical model arises from the purported random mathematical structure requires us to take the most general, random mathematical structure $\mathcal{M}$.  Here again a graph is a very general structure linking with an edge only those points which are related and model theory provides a very general formalism and body of results giving substructures and in turn through their properties, primitive laws~\cite{tarzi1}.  

The various number systems can be arrived at from the mathematical structures.  So for example natural numbers count the number of substructures and subtracting the number of certain types of substructures from the set of other types gives the integers.  Integers also count the number of vertices in $\mathfrak{R}$, whilst being of uncountable order the real numbers enumerate the number of elements of the group $\Aut(\mathfrak{R})$~\cite{tarzi1}.
\index{group ! Aut(\mathfrak{R})@$\Aut(\mathfrak{R})$}%

\medskip

Eventually a manifold structure will be required (see below), for which real numbers are necessary.  The uncountable order of the automorphism
\index{group ! automorphism}%
 group $\Aut(\mathfrak{R}_{m, \omega})$
\index{group ! Aut(\mathfrak{R}_{m, \omega})@$\Aut(\mathfrak{R}_{m, \omega})$}%
 of the multicoloured random graph
\index{graph ! random ! $m$-coloured}%
 on $m$ colours is one source.  Complex numbers can arise from a definition of an action principle in physics with a complex exponent.   The geometric meaning of the imaginary ``i'' could simply be the introduction of rotations, as its basic action on the complex plane would attest.  We emphasized the fundamental nature of the rotation group in~\cite{tarzi2}.  
\index{group ! orthogonal (rotation)}%

Physics at energies less than 300 GeV is called `infrared' and above 1 TeV is called `ultraviolet'.   In the infrared limit an asymmetric $\lambda_{ijkl} \phi_i \phi_j \phi_k \phi_l$ theory, under suitable conditions, becomes $O(n)$-invariant.
\index{group ! orthogonal (rotation)}%

\smallskip

There are many topologies and distance functions that are derivable from the random graphs.
\index{graph ! random ($\mathfrak{R}$)}%
 Distance within a relational structure
\index{structure ! relational}%
 $\mathcal{M}$ may be a function of the degree of difficulty in going from one substructure to another.  Random graphs have a \emph{topology of pointwise convergence}
\index{topology ! of pointwise convergence}%
 of permutation groups
\index{group ! permutation}%
 that they support.  A \emph{permutation group} acting on a set $X$ is a subgroup of the symmetric group $\Sym(X)$.
\index{group ! symmetric ($\Sym$)}%
 The action is \emph{transitive}
\index{group ! permutation ! transitive}%
if the only fixed subsets of $X$ are the empty set $\emptyset$ and the entire set $X$,
and \emph{primitive}
\index{group ! permutation ! primitive}%
if in addition the only fixed partitions of $X$ are $\{X\}$ and the partition into singletons.  Thus a group is \emph{primitive} if there are only the two trivial congruences, which are the relation of equality and the
`universal' relation with $x \sim y$ for all $x, y \in X$.    A group is \emph{$n$-transitive}
\index{group ! permutation ! $n$-transitive}%
if it is transitive on the set of ordered $k$-subsets of $X$ and it is \emph{$n$-homogeneous}
\index{group ! permutation ! $n$-homogeneous}%
if it is transitive on the set of unordered $n$-subsets of $X$.

Let $G$ be a permutation group
\index{group ! permutation}%
 acting on a countably infinite set $X$.  The natural topology
\index{topology}%
  on permutation groups
\index{group ! permutation}%
 is the \emph{topology of pointwise convergence}
\index{topology ! of pointwise convergence}%
in which for a sequence $g_n \in G$ of permutations $G= \Sym(X)$ on $X$, $\lim_{n \to \infty} (g_n) = g \in G$ if and only if $\forall x_i \in X,\ \exists n_0 \in \omega$ such that $\forall n > n_0,\ x_i g_n = x_i g$.  (We shall assume that the permutation groups
\index{group ! permutation}%
 have countable degree; for arbitrary degree we would have to consider limits of nets rather than sequences.) 

Endowing a group $G$ with this topology
\index{topology}%
 turns it into a \emph{topological group},
\index{group ! topological}%
that is multiplication and inversion are continuous, so that if $g_n \to g$ and $h_n \to h$ then $g_n h_n \to gh$ and $g_n^{-1} \to g^{-1}$.  Further endowing $G$ with a metric turns it into a complete metric space.
\index{metric space ! complete}%

This in turn can potentially give a manifold-like structure.

It maybe that the graph-theoretic content of manifolds will eventually have to be considered.  A lucid introduction to both the theory of graphs in surfaces and to homology on graphs is~\cite{giblin} but also see the more recent~\cite{ellis}.

\smallskip

One of the few assumptions permitted in the Nielsen
\index{Nielsen, H. B.}%
 program is the existence of \emph{exchange forces} which permute subsystems of the abstract set of points  thereby transposing particle states.  That is, in a pair of particle states, the exchange force will exchange the two different particles occupying these states with each other.  There are a large variety of possible types of permutations of random graph
\index{graph ! random ($\mathfrak{R}$)}%
 vertices that may be brought to bear here.  We mentioned possible uses of \emph{switching automorphisms} 
\index{group ! switching automorphism}%
and \emph{almost automorphisms}
\index{group ! almost automorphism}%
 in modeling physics in~\cite{tarzi2}.  A permutation $g \in \Sym(X)$ of $X$ is an \emph{almost automorphism}
\index{group ! almost automorphism}%
of a multicoloured graph $\Gamma$ on a vertex set $X$ if the colour of $\{xg,yg\}$ is equal to the colour of $\{x,y\}$ for all but finitely many $2$-element subsets of $X$. That is a permutation of the vertex set of $\mathfrak{R}_{m,\omega}$ is an almost automorphism if the set of edges whose colour it alters is finite.   We discuss switchings in a later chapter.

Just as force is a gradient of energy with respect to position coordinates, so exchange forces can generically be understood as the gradient with respect to the difference in energy depending on whether or not the Pauli exclusion principle
\index{exclusion principle}%
 applies.  Exchange forces are central to the emergence of bosons and fermions, and so in rendering particles that are all different at fundamental scale but are effectively identical at low energies.   One possible symmetry is a \emph{permutation}
\index{group ! permutation}%
 of two sets of particles which corresponds to a boson symmetry between the two sets.  This could be a local automorphism
\index{automorphism}%
 of two subsets of graph vertices.  Perhaps the permutation of two sets of fermions could be effected by anti-automorphisms.   An \emph{anti-automorphism}
\index{graph ! anti-automorphism}%
of a graph $\Gamma$ is an isomorphism from $\Gamma$ to the complementary graph $\overline{\Gamma}$ wherein edges and non-edges are interchanged.  The idea is that the particles are so restricted that they behave to all intents and purposes as if they were identical in almost all respects, and in this way they become subject to permutation symmetry operations.

Nielsen
\index{Nielsen, H. B.}%
 argues in his blog post that the addition of an extra particle to an already huge set of particles could be an argument for the $\SU(2)$
\index{group ! SU(2)@$\SU(2)$}%
 symmetry for rotation
\index{group ! orthogonal (rotation)}%
 of the superposition of the set of particles and thereby the particle should effectively have an $\SU(2)$ symmetry.  The random graph
\index{graph ! random ($\mathfrak{R}$)}%
 has stability properties such as for each vertex $v \in \mathfrak{R}$, $\mathfrak{R} \cong \mathfrak{R} \backslash \{v\} \cong \mathfrak{R} \cup \{v\}$.

\smallskip

The starting point of the Random Dynamics project
\index{Random Dynamics}%
 should be the consideration of as large a class of conceivable `theories' or `models' as possible.  A probability measure might then be chosen for the set of models.  The hypothesis is that a model chosen at random from this set will almost always contain the empirically known laws of nature; that is all the presently known laws of physics will follow from almost any model which is sufficiently complicated enough, provided we go to some generic low energy limit.

Froggatt and Nielsen
\index{Froggatt, C. D.}%
\index{Nielsen, H. B.}%
 state~\cite[p.~139]{froggatt}:

``At first, it may seem rather hopeless to try to derive low energy physics, or even the quantum field theory of glass, out of a totally random mathematical model.  However we believe some concepts are so general that we can hope to find them relevant in almost any sufficiently rich and complicated `model'.  For example it should be possible to introduce a crude concept of distance, or some sort of topology,
\index{topology}%
 on almost any mathematical structure $S$ which contains a huge number of `elements' having relations between themselves: the longer the chains of intermediate elements needed to establish a relation between two `elements', the further apart the two `elements' are defined to be.  For such a mathematical structure, rich in the sense of having a huge number of elements and also being rather repetitive, it must be possible to define the concept of small modifications $[\ldots]$ of the structure $S$.''

Multicoloured random graphs
\index{graph ! random ! $m$-coloured}%
 would then seem to fit this approach and to provide us with a mathematical formalism that produces meaningful theory.

So the idea behind the Random Dynamics
\index{Random Dynamics}%
 program is that from almost all choices of the a priori fundamental laws based on structural stability,  the resulting physical laws are unavoidable~\cite[p.~566]{froggatt}, so that a small variation of the basic dynamics that is of the equations of motion, should not lead to an essential change. 

There is one important example that rhymes with this idea.  
Poincar\'e invariance
\index{group ! Poincar\'e}%
 is a given subgroup of the diffeomorphism reparameterisation group
\index{group ! diffeomorphism}%
 of general relativity, which under the transformation
\[ x^{\mu} \to x^{'\mu} = \Lambda^{\mu}_{\nu} x^{\nu} + a^{\mu}. \]
relates two coordinate systems by a Lorentz transformation $\Lambda^{\mu}_{\nu}$ followed by a translation $a^{\mu}$.
  As we mentioned in~\cite{tarzi2}, one reason for the fundamental importance of the Poincar{\'e} group
\index{group ! Poincar\'e}%
 is that the Poincar{\'e}-invariant linear wave equations that are based on special relativistic quantum theory (those of Klein \& Gordon,
\index{Klein-Gordon equation}%
 Weyl,
\index{Weyl equation}%
 Dirac,
\index{Dirac equation}%
 Maxwell,
\index{Maxwell's equations}%
 Proca,
\index{Proca equation}%
 Rarita \& Schwinger,
\index{Rarita-Schwinger equation}%
 Bargmann
\index{Bargmann-Wigner equation}%
\index{Bargmann, V.}%
    \& Wigner,
\index{Wigner, E. P.}%
  Pauli \& Fierz)
\index{Pauli-Fierz equation}%
are projections of an irreducible sub-representation
\index{irreducible representations}%
   of the (universal cover $\mathbb{R}^4 \sd SL(2, \mathbb{C})$ of the) Poincar{\'e} group~\cite{giulini}.
\index{group ! Poincar\'e}%
\index{group ! SL(2, \mathbb{C})@$\SL(2, \mathbb{C})$}%
   Perhaps then the Poincar{\'e} group
\index{group ! Poincar\'e}%
  may be considered as being prior to both matter and spacetime.   Furthermore, physical quantities such as energy, momentum, and angular momentum can be \emph{defined} as the conserved quantities associated to spacetime automorphisms
 \index{automorphism}%
 via Noether's Theorem.
\index{Noether's Theorem}%

Within the Random Dynamics
\index{Random Dynamics}%
 program itself,
Froggatt and Nielsen
\index{Froggatt, C. D.}%
\index{Nielsen, H. B.}%
 study a very general quantum field theory (QFT), which is not even assumed to be Lorentz invariant,
\index{Lorentz invariance}%
 in the limit of free and lowest energy approximation~\cite{froggattaa}.  A relativistic theory with just three \index{Weyl equation}%
 and in the boson case in the form of the Maxwell equations.
\index{Maxwell's equations}%
  However this works for one particle species on its own and does not, immediately at least, lead to Lorentz invariance
\index{Lorentz invariance}%
 if many particle species are involved.  The three space dimensions are derived from the fact that whilst the basic model assumes an arbitrary number of dimensions and has momentum degrees of freedom in all these dimensions, the velocity components in all but three dimensions turn out to be zero.

Thus almost all equations of motion would lead to internal symmetries such as parity, isospin, etc., relativistic invariance and other special principles.  Then if known physics emerges as a consequence of all but those belonging to a special null measure set then there would be no need to consider laws of nature at all or at least only as approximations in some limit, and that the parameters of the fundamental dynamics are general rather than of a special type.  Relativity and 3-dimensionality of space could be derived in the limit of low energy per particle.

More formally begin with the largest conceivable class of models and choose a measure on this class.  Using a probability measure we could ask for the probability that a given class of models is Lorentz invariant.
\index{Lorentz invariance}%
  It could also lead to symmetries in certain limits~\cite[Section 2, Chapter 6]{tarzi2}.

The hope is that neither the choice of measure nor the choice of parameters used to specify a measure on all the lattice
\index{lattice}%
 models is significant~\cite[p.~571]{froggatt}.  The search would be for generic properties, derived from the topology
\index{topology}%
 on the set of objects or theories being studied.  Such properties would almost certainly be true when a measure exists on the set.  

A subset of a topological space is \emph{dense}
\index{topological space ! dense}%
if it meets every nonempty open set.
\index{set ! open}%
A space admitting a countable dense set
\index{set ! dense}%
is called \emph{separable}.
\index{topological space ! separable}%
  A \emph{Polish space}
\index{Polish space}%
is a topological space that is separable and completely metrizable,
\index{topological space ! completely metrizable}%
meaning that its topology
\index{topology}%
 is induced by a complete metric.  A topological group is \emph{Polish}
\index{group ! topological ! Polish}%
if the underlying topology of the group is that of a Polish space, with the open subgroups forming a base of open neighbourhoods of the identity $1$.  

A property $\mathcal{P}$ of elements of a topological
\index{topological space}%
 space $\mathcal{S}$ is said to be \emph{generic}
\index{structure ! generic}%
 if the set $\{X \in \mathcal{S}\ |\ \mathcal{P} \text{ is true for}\ X \}$ is the intersection of countably many dense open subsets
\index{set ! dense open}%
 of $\mathcal{S}$.  For example, being irrational is a generic property of the real numbers.  The notion of genericity is standard and already well-studied when considering homogeneous structures
\index{structure ! homogeneous}%
 such as random graphs.
\index{graph ! random ($\mathfrak{R}$)}%

One way to construct the countable random graph
\index{graph ! random ($\mathfrak{R}$)}%
 that is equivalent to the usual method is
as a countable \emph{generic structure}
\index{structure ! generic}%
where generic means construction by finite approximations.  Random graphs
\index{graph ! random ($\mathfrak{R}$)}%
 are defined in a space of graphs with a given property.  The potential difficulties with this process are firstly that of defining a measure
\index{measure}%
on this space and secondly that of determining whether or not a limit exists.  Even if a limiting structure exists it could be a surprising one.  That a random structure is not always equivalent to a generic structure
\index{structure ! generic}%
 is exemplified by triangle-free graphs~\cite{tarzi1}.
\index{graph ! triangle-free}%

Alternatively a \emph{generic structure} 
\index{structure ! generic}%
is one that is residual in some natural complete metric space in the sense of the \emph{Baire category theorem}
\index{Baire category theorem}%
 (see Category and Measure Appendix in~\cite{tarzi1}).  This says that we can put a natural metric structure on the class of all objects with a given \emph{age} or smaller so that the isomorphism class of such objects is residual, or in other words that almost all objects look like the one we are interested in.   

Endowing a topological group $G$ with a metric turns it into a complete metric space, so we can use the Baire category theoretic
\index{Baire category theorem}%
notion of \emph{meagre}.
\index{meagre set}%
  The topology
\index{topology}%
 is generated by basic open sets
\index{set ! basic open}%
 of the form $[p] = \{g \in G : g\ \text{agrees with}\ p\ \text{on its domain} \}$ where $p$ is a $1$--$1$ map from a finite subset of $X$ to $X$.  The automorphism group
\index{group ! automorphism}%
 of any relational structure
\index{structure ! relational}%
  is a closed
\index{group ! closed}%
 subgroup of $\Sym(X)$ and so is automatically a complete metric space, and therefore we can talk about generic automorphisms.  A \emph{generic automorphism} 
\index{automorphism ! generic}%
 is one whose conjugacy class within the group is large in the sense of Baire category.  One way of capturing the notion of $g$ being a `typical' element is to require $g$ to lie in certain \emph{dense open sets}.  
\index{set ! dense open}%
An example of a dense open set for any $g \in G$ is the set $\{G \backslash g\}$. 

A homogeneous structure
\index{structure ! homogeneous}%
 will have a generic automorphism, where a \emph{generic element}
\index{automorphism ! generic}%
of the automorphism group
\index{group ! automorphism}%
 is one that lies in a comeagre conjugacy class relative to the given topology.
\index{topology}%
  Let $G$ be a permutation group
\index{group ! permutation}%
 acting on a countably infinite set $X$.  

For more  on the notion of \emph{generic automorphism}
\index{automorphism ! generic}%
 see~\cite{tarzi1}.

\bigskip

In mathematics, a lattice is a vector space.  More accurately, a lattice $\mathbb{L}$~\label{mathbb{L}}
\index{lattice}%
 in a real finite-dimensional vector space $V$ is a subgroup of $V$ satisfying one of the following equivalent conditions:

(i)  $\mathbb{L}$ is discrete and $V / \mathbb{L}$ is compact;
\index{group ! discrete}%
\index{group ! compact}%

(ii)  $\mathbb{L}$ is discrete and generates the $\mathbb{R}$-vector space $V$;

(iii)  $V$ has an $\mathbb{R}$-basis $(e_1, \ldots, e_n)$ which is a $\mathbb{Z}$-basis of $\mathbb{L}$ (that is $\mathbb{L} = \mathbb{Z} e_1 \oplus \ldots \oplus \mathbb{Z} e_n)$.

So a lattice in $\mathbb{R}^{n}$ is the set of integral linear combinations of $n$ linearly independent vectors.  The lattices generated by vectors of norm $1$ are $\mathbb{Z}^n$ with the standard bilinear form.

\bigskip

In this brief book we can only give a flavour of the random graph
\index{graph ! random ($\mathfrak{R}$)}%
 mathematics that can be put to use in modelling Random Dynamics.
\index{Random Dynamics}%
One question we could seek to answer is: given a certain natural topology
\index{topology}%
 on the set of all lattice
\index{lattice gauge theory (LGT)}%
 theories show that Lorentz invariance
\index{Lorentz invariance}%
 is a generic property in the low energy limit.  A lattice
\index{lattice}%
 could reasonably be taken to be a low energy limit of a graph.

\bigskip

Amongst the derived symmetries to which it is hoped that the Random Dynamics approach can be applied~\cite[Chapter~VII]{froggatt} are:

(i) macroscopic scaling symmetry present for a very large number of molecules but absent at the atomic scale;

(ii) five symmetries derived from the SM, (a) charge conjugation invariance, (b) parity invariance, (c) time reversal invariance, (d) flavour conservation (including any baryon and lepton number conservation), and (e) chiral symmetry  (including Gell-Mann-Ne'eman $\SU(3)$).
\index{group ! SU(3)@$\SU(3)$}%
\index{Gell-Mann, M.}%
\index{Ne'eman, Y.}%

There are exact symmetries for example $\mathcal{CPT}$ invariance, baryon number conservation and conservation of lepton number for each generation.  Approximate symmetries such as $\mathcal{C}$, $\mathcal{P}$, and $\mathcal{T}$,
\index{Ch@charge conjugation symmetry ($\mathcal{C})$}%
\index{Parity@Parity symmetry ($\mathcal{P})$}%
\index{Time@time reversal symmetry ($\mathcal{T})$}%
 flavour and chiral symmetries are derived.

Given that parity and strangeness conservation are derivable from the strong and electromagnetic interactions there is no reason why weak interactions should conserve them.

Poincar\'e symmetry,
\index{group ! Poincar\'e}%
 including rotational
\index{group ! orthogonal (rotation)}%
 and space-time translational symmetries are considered part of the diffeomorphism
\index{diffeomorphism}%
 symmetry of general relativity.  So a proper account of their origin must go beyond known theory to consideration of say a pregeometric theory, as must any account of gauge invariance.
\index{gauge ! invariance}%

But it is not merely symmetries whose origin random dynamics aims to uncover.  The idea goes further in suggesting that almost any theory proposed at random will in one or more limits, say at low energies, possess the regularity in question.  Almost any model at the fundamental level would then be a theory of everything rather than one specific one.  The observed physical laws would seem to be stable under changes of the fundamental model.  This random dynamics philosophy overlaps with that of Wheeler's `Law without Law'~\cite{wheeler}.

\chapter{Emergence of Space and Quantum Mechanics}

The following discussion~\cite{froggatt} is preliminary and abstract.  Beginning with a random mathematical structure $\mathcal{M}$ space-time must be introduced and physical objects must be identified.

Physical states or `wavefunctions' are defined on a topological space
\index{topological space}%
 $S_{\mathcal{M}}$ constructed from the structure.  A topology
\index{topology}%
 on the random structure would give the concept of distance.  

The vector space of small modifications of $\mathcal{M}$ is identified with the Hilbert space of a generalised QM, and the topological space on $\mathcal{M}$ is the configuration space of functions for a quantum field-type theory, that is, each point in the topological space corresponds to a field configuration.  The field values 
over space-time become the coordinates for $S_{\mathcal{M}}$ which then essentially becomes a differentiable manifold.  Then assume that it is possible to find a tangent vector space to $S_{\mathcal{M}}$.

Introduce a topological structure on the basis vectors or `directions' in the tangent space, so as to identify coordinates in $S_{\mathcal{M}}$ with points or small regions in a pregeometrical space.  These directions could form another complicated mathematical structure and again as for $S_{\mathcal{M}}$, two `directions' are close together if a simple mathematical operation connects them.

Groups of `directions' in the tangent space can form neighbourhoods or small overlapping structures associated with small regions in a pregeometrical space-time.  Associating field values, that is points of $S_{\mathcal{M}}$, with small regions of space-time ensures a local field theory in the long wavelength limit.

Next identify a certain configuration as the vacuum state
\index{vacuum}%
 and a symmetry group of gauge transformations
\index{gauge ! transformation}%
\index{group ! gauge}%
 including ones that might form a diffeomorphism group
\index{group ! diffeomorphism}%
 permuting points of the pregeometric space which may themselves be identified with space-time points.  Orbits under diffeomorphism group action could then form space-time manifolds.

\medskip

In Random Dynamics
\index{Random Dynamics}%
 the action is assumed to be that of a discretized lattice-type QFT
\index{lattice}%
 rather than the usual continuum theory.  The founders chose the description \emph{glass} by analogy with the ``frozen in'' structure of a spin glass.  The resulting QFT glass is modelled as a Feynman path integral
\index{path integral}%
 with constant (or in the terminology \emph{quenched}) random parameters.  Then quantum fluctuations are thought to smooth out the QFT glass into a continuum theory.   

In building a field theory glass model, a generalised quantum field $\phi(i)$ is defined on a set of quenched random space-time points $\{ i \}$ and takes values in quenched random manifolds $M_i$. Quenched parameters are chosen randomly but are then fixed in the path integral.  The field theory glass action
\[ \mathcal{S}[\phi] = \sum_r \mathcal{S}_r(\phi(i)),\quad i \in r, \]
is a sum over contributions $\mathcal{S}_r(\phi(i))$ from many small overlapping regions $r$ in space-time.  A gauge theory
\index{gauge ! theory}%
 for which the gauge symmetry group
\index{group ! gauge}%
 varies randomly across the regions, called a \emph{gauge glass},
\index{gauge ! glass}%
 is constructed on which continuum gauge field degrees of freedom $A_{\mu}^{a} (x)$ can be imposed.  A \emph{gauge glass}
\index{gauge ! glass}%
 model has randomly varying parameters on the ``lattice'', though the same random value at any given plaquette is retained throughout; the value is then called \emph{quenched}.  

We take up the issue of the gauge group having nontrivial outer automorphisms
\index{automorphism ! outer}%
 in Chapter~\ref{oaut}.

\bigskip

Let $N_{\mathcal{M}}(X)$ be the number of substructures $X$ of the relational structure
\index{structure ! relational}%
 $\mathcal{M}$ and $N_{X}(Y)$ be the number of substructures $Y$ of $X$.  Clearly, these are natural numbers including zero.

A coordinate system can be formed in which we can associate one substructure with each coordinate.  The whole of $\mathcal{M}$ would then be describable by a vector in a vector space with a basis in corespondence with the substructures. 

A mathematical structure such as a graph is a \emph{Cayley object},
\index{Cayley object}%
$O$, for the group $G$ if its point set are the elements of $G$ and right multiplication by any element of $G$ is an automorphism of $O$.  A Cayley graph
\index{graph ! Cayley}%
for a group $G$ takes the form $\Cay(G, S)$, where $S$ is an inverse-closed subset of $G \backslash \{1\}$ with vertex set $G$ and edge set $\{\{g, sg\}: g \in G, s \in S\}$.  As $G$ is countable, we can enumerate the inverse pairs of non-identity elements of $G$ as $\{g_{1}, g_{1}^{-1}\}, \{g_{2}, g_{2}^{-1}\}, \ldots$.  Baire
category theory
\index{Baire category theorem}%
 can sometimes be used for building a homogeneous Cayley object
\index{Cayley object}%
for a group $G$, by showing that almost all $G$-invariant objects of the required type, that is a residual set of them, are homogeneous.
\index{structure ! homogeneous}%
  In order to make sense of the notion of a residual set of Cayley graphs for $G$, specify Cayley graphs by paths in a ternary tree, whereby the three descendents of a node or vertex at level $n$ correspond to including or excluding the inverse pair $\{g_{n+1}, g_{n+1}^{-1}\}$ in one of the colour classes of $G$; call these colour classes red $(\mathfrak{r})$, blue $(\mathfrak{b})$ and green $(\mathfrak{g})$.
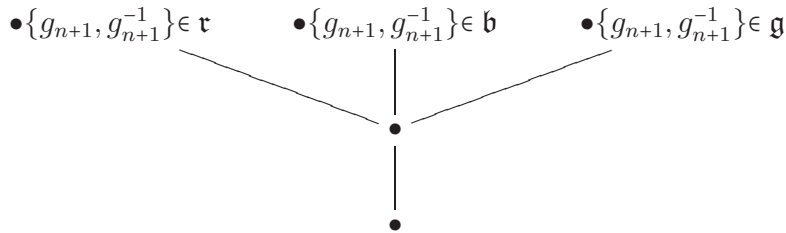
\begin{figure}[!h]
$$\xymatrix{
& {\bullet}{\{g_{n+1}, g_{n+1}^{-1}\}} {\in \mathfrak{r} } \ar@{-}[dr] & {\bullet} {\{g_{n+1},
g_{n+1}^{-1}\}} {\in \mathfrak{b} } \ar@{-}[d] & {\bullet}{\{g_{n+1}, g_{n+1}^{-1}\}} {\in \mathfrak{g} }
\ar@{-}[dl]\\
&& {\bullet} \ar@{-}[d]\\
&& {\bullet}
}$$\caption{Colour classes of $G$}
\label{ccif}
\end{figure} 

The set of three-coloured Cayley graphs is identified with the set of paths in the ternary tree.  In Baire category theory
\index{Baire category theorem}%
 if an object is specified by a
countable sequence of choices, then the existence of one such object
with a given property $\mathcal{P}$ can be proved by showing that
$\mathcal{P}$ holds for `almost all' choices.  Let $\mathcal{P}(T)$
denote the set of paths of countable length starting at the root of a
tree $T$, whose nodes at height $n$ are labelled by structures on
$\{1, \ldots, n\}$.  We define the distance between distinct paths $p$
and $p'$ to be $f(n)$, where $n$ is the height of the last node at
which $p$ and $p'$ agree, and $f$ is any strictly decreasing function
tending to zero.  The complete metric space 
\index{metric space ! complete}%
 to which the Baire
category theorem will be applied arises from paths in rooted trees of
countable height.  A Cauchy sequence
\index{Cauchy sequence}%
in this space is a sequence of
paths agreeing on increasingly longer initial segments, and so has a
unique limiting path.  In this way a complete metric space can be built.  

We require an interpretation of openness and denseness
\index{set ! dense open}%
 to formulate residual sets in this space.  An \emph{open ball}
\index{open ball}%
consists of all paths in the tree containing a given vertex.  A set $S$ of paths is open
\index{set ! open}%
 (or \emph{finitely determined}) if each path in $S$ has a vertex such that
every path through this vertex is in $S$.  A set $S$ is \emph{dense}
(or \emph{always reachable}) if it meets every open ball, i.e. if all
vertices lie on some path in $S$.  The triality graph
\index{graph ! triality ($\mathfrak{R^{t}}$)}%
is the Fra\"{\i}ss\'e limit
\index{Fra\"{\i}ss\'e limit}%
of the class of all appropriately
defined finite $3$-coloured graphs, so with a countable vertex set the
isomorphism class of $\mathfrak{R^{t}}$
\index{graph ! triality ($\mathfrak{R^{t}}$)}%
 is residual in the set of $3$-coloured graphs on vertex set $\mathbb{N}$.  Thus it makes sense to talk of a residual set of Cayley graphs for $G$.  The metric space has an underlying topological space,
\index{topological space}%
 where the topology
\index{topology}%
 lies on the collection of inverse closed subsets of $G$.

\bigskip
\bigskip

For a physical interpretation we may be able to identify the types of substructures with possible paths in the Feynman path integral.
\index{path integral}%

That is to say $e^{i/\hbar} \mathcal{S}(path) = N_{\mathcal{M}}(X(path))$, where $X(path)$ is the substructure corresponding to \emph{path}. 

Another possibility is that the path from one vertex to another in the graph is what identifies the discrete parts of a path integral.

The transition amplitude from an initial to final state in usual Feynman path integral is the funtional integral
\[ < f | U(t_i, t_j) | i > = \int e^{\frac{i}{\hbar} \mathcal{S}_{t_i t_f} (path)}\ \mathcal{D}path, \]
where $\mathcal{S}_{t_i t_f} (path)$ is the integral of the Lagrangian is over the path being integrated in between times $t_i$ and $t_j$.  

The fields remain real or complex as usual but the coupling constants and squared masses are assumed complex giving $\mathcal{S}(history) = \mathcal{S}_R(history) + i \mathcal{S}_I (history)$.  

Adding an imaginary part to the Lagrangian density allows the action to be complex.  The imaginary part $\mathcal{S}_I(path)$ of the action is a weight function of the usual real-action Feynman path integral
\index{path integral}%
 and only when $\mathcal{S}_I(path)$ is very negative will the factor $e^{-\mathcal{S}(path)}$ dominate the Feynman path integral.  An action with both real and imaginary parts leads to the second law of thermodynamics.  

The imaginary part $\mathcal{S}_I$ of the action can be thought of as the time-integral over the imaginary part of the Lagrangian for $t \in (- \infty, \infty)$.  One consequence of introducing an imaginary part to the action used in the path integral,
\index{path integral}%
 of a similar form to the usual real part but with different parameters, is a model determining not only equations of motion but also the initial conditions by typically providing a large factor in the probability thereby fixing the path obeyed by the equations of motion and the initial conditions.  The imaginary part $\mathcal{S}_I$ of $\mathcal{S} = \mathcal{S}_R + \mathcal{S}_I$ fixes a state in both the future and past, for example the particle's position when it went past the double slit screen.  

Two approaches to extracting probabilities and expectation values are then outlined~\cite{nielsenninomiya}.  Firstly allowing a natural definition of an average of quantity $\mathcal{O}$ as a function of the fields $\phi$
\[ \langle \mathcal{O}(\phi)\rangle = \frac{\int e^{i\mathcal{S}[\phi]} \mathcal{O}(\phi) \mathcal{D} \phi}{\int e^{i\mathcal{S}[\phi]} \mathcal{D} \phi} \]
may give a complex probability in a path integral.  The second approach inserts a series of projections onto small regions for operators and take the modulus square of the integrals between initial and final values.  Under certain assumptions the two approaches should approximately agree with each other.

It is hoped that the Feynman path integral formulation could arise from Random Dynamics
\index{Random Dynamics}%
 plus a few extra assumptions~\cite{bennettkleppe}, and it seems that taking an imaginary action is what is required~\cite{nielsenhb} because this determines which of the solutions of the equations of motion is the right one and thus determines initial conditions.

\bigskip

In~\cite{bennettkleppe} Bennett, Kleppe and Nielsen
\index{Bennett, D. L.}%
\index{Kleppe, A.}%
\index{Nielsen, H. B.}%
  discuss the possibility of random complicated structures leading to manifolds, via concepts of connectivity and nearness or resemblance of one substructure with another.  The
numbers of substructures determining the coordinates of the mathematical structure are given by a matrix which shows how to get to certain substructures from others.  In particular, with as many rows and columns as there are number of types of substructures, that is paths.  

Matrices that grow to reflect subgraphs on the way to constructing the random graph
\index{graph ! random ($\mathfrak{R}$)}%
 have also been used, as we shall now see.  The edge-set of a simple graph on a given countable vertex set can be thought of as a countable zero-one sequence,
\index{zero-one sequence}%
  leading to the fact that the class of graphs forms a complete metric space, of which the subclass of graphs satisfying the two-colour injection property
\index{injection property}%
 (equivalent to the \emph{one-point extension property} $(*_2)$ for $2$ edge-coloured graphs) is a residual subset~\cite{tarzi1}.

The metric space equivalent of the random graph
\index{graph ! random ($\mathfrak{R}$)}%
 is uniquely defined by
a one-point extension property
\index{one-point extension property}%
 as it is for random graphs.   A metric space $X$ is called a \emph{(generalized) Urysohn
space}~\cite{urysohn}
\index{Urysohn space}%
if whenever $A \subseteq X$ is a finite metric subspace of $X$
and $A' = A \cup \{a\}$ is an arbitrary one-point metric space
extension of $A$, the embedding $A \hookrightarrow X$ extends to an
isometric embedding $A' \hookrightarrow X$.  Up to isometry there is
only one universal complete separable Urysohn metric space, denote it
$\mathfrak{U}$, which contains an isometric copy of every separable
metric space. 

The way to show that $\mathfrak{U}$ is the `Random Polish Space'
\index{Polish space}%
is to choose $n$-tuples of random real numbers as distances from $n$-dimensional
Euclidean space
\index{Euclidean space}%
 $\mathbb{R}^n$, take a measure
\index{measure}%
for each dimension, and construct a
countable space one point at a time using a suitable measure on
extensions via a `random metric' between the $(n+1)^{\text{st}}$ point and
the first $n$ points, ensuring that the triangle inequality which determines a cone in $\mathbb{R}^n$ is satisfied each time.  Finally take the completion of this.  We can say more precisely what a random metric on $\mathbb{N}$ looks like.  Let $a_0, a_1$ be the first two points of the space.  Choose $d(a_0, a_1) = x_0^{(1)} \ge 0$, to lie in $\mathbb{R}^{+}$, where the superscript denotes dimension.  Next, choose $a_2$ so that  $d(a_0, a_2) = x_0^{(2)} = x \ge 0$ and $d(a_1, a_2) = x_1^{(2)} = y \ge 0$ satisfying
\[|x - y| \le x_0^{(1)} \le x + y.\]   Now choose $a_3$ so that  $d(a_0, a_3) = x_0^{(3)} = z_1 \ge 0$, $d(a_1, a_3) = x_1^{(3)} = z_2 \ge 0$ and $d(a_2, a_3) = x_2^{(3)} = z_3 \ge 0$ satisfying several inequalities of the form
\[|z_1 - z_2| \le x_0^{(2)} \le z_1 + z_2.\]  
Each cone at each new dimension depends on the previous one.  By choosing a wide range of reasonable measures on all cones, the completion of $\mathbb{N}$ equipped with a random metric is almost always isometric to $\mathfrak{U}$.  A. Vershik~\cite{vershik}
\index{Vershik, A. M.}%
attaches admissible vectors to distance matrices in order to mimic the one-point extension property.
\index{one-point extension property}%
  The indivisibility property of $\mathfrak{U}$ whereby if $B \subset \mathfrak{U}$ is an open ball
\index{open ball}%
then $\mathfrak{U} \backslash B \cong \mathfrak{U}$ is equivalent to removing a row and some columns in the distance matrices stabilizing $\mathfrak{U}$.

It is possible to construct $\mathfrak{R}$ from $\mathfrak{U}$. Any group acting on $\mathfrak{U}$ with a countable dense orbit preserves the structure of $\mathfrak{R}$ on the dense orbit.  

\bigskip

Topology
\index{topology}%
 is in a sense a generalisation of geometry where sets of points and their relation to each other is the central idea rather than studying solid objects.  There is a natural topology
\index{topology}%
 for the ordinary two-colour random graph
\index{graph ! random ($\mathfrak{R}$)}%
 which may well have a physical interpretation, and that is the \emph{neighbourhood topology}~\cite{tarzi1}.
\index{topology ! neighbourhood}%

In~\cite[Chapter~8]{tarzi1} we studied a filter,
\index{filter}%
 to be defined shortly, and a topology defined naturally from $\mathfrak{R}$, generated by the vertex neighbourhoods in $\mathfrak{R}$, and whose automorphism groups
\index{group ! automorphism}%
 contain the automorphism group of $\mathfrak{R}$. For such neighbourhood filters, $\mathfrak{R}$ has a `universal' property: any countable
graph whose neighbourhood filter
\index{filter ! neighbourhood}%
 is non-trivial contains $\mathfrak{R}$ as a spanning subgraph.
\index{graph ! spanning}%
  From the injection property of $\mathfrak{R}$ it follows that any finite vertex set of $\mathfrak{R}$ has a common neighbour, and this property characterizes the class of countable graphs containing $\mathfrak{R}$ as a spanning subgraph.
\index{graph ! spanning}%

A \emph{filter}
\index{filter}%
 on a set is a family $ \mathscr{F}$ of subsets of the vertex set $V(\mathfrak{R})$ of $\mathfrak{R}$
\index{graph ! random ($\mathfrak{R}$)}%
 satisfying
\begin{itemize}
\item $X,Y\in\mathscr{F}$ implies $X\cap Y\in \mathscr{F}$;
\item $X\in \mathscr{F}$, $Y\supseteq X$ implies $Y\in\mathscr{F}$ (upward closed).
\end{itemize}

A filter $\mathscr{F}$ is \emph{trivial}
\index{filter ! trivial}%
 if it consists of all subsets of $V(\mathfrak{R})$; it is
\emph{principal}
\index{filter ! principal}%
 (one-generated) if it consists of all sets containing a fixed subset $A$ of
$\mathscr{F}$; and it is an \emph{ultrafilter}
\index{filter ! ultrafilter}%
\index{ultrafilter}%
 if, for any $X\subseteq V(\mathfrak{R})$, just one of $X$ and $V(\mathfrak{R})\setminus X$ belongs to $\mathscr{F}$. 

\medskip

If $\mathscr{F}$ is a filter,
\index{filter}%
 then $\mathscr{F}\cup\{\emptyset\}$ is a topology.
\index{topology}%
  Since $V(\mathfrak{R})$ is countable, no non-discrete metric on $V(\mathfrak{R})$ can be complete.  If the topology is not discrete then the set of complements of finite unions of discrete sets is a filter. 
\index{filter}%

Let $\Gamma$ denote a graph and $\Gamma(v)$ be the set of vertices attached to vertex $v$ in the graph.

\begin{proposition}
Suppose that $\Gamma$ has the property that each vertex has a non-neighbour.
Then the filter
\index{filter}%
 generated by the closed neighbourhoods
$\overline{\Gamma}(v)=\Gamma(v)\cup\{v\}$ is equal to $\mathscr{F}_{\Gamma}$.
\end{proposition}

Let $\mathfrak{R}$ denote the countable random graph.
\index{graph ! random ($\mathfrak{R}$)}%

\begin{proposition}
\label{prop8.2}
The following three conditions on a graph $\Gamma$ are equivalent:
\begin{itemize}
\item[(a)] $\mathscr{F}_\Gamma$ is nontrivial;
\item[(b)] $\Gamma$ contains $\mathfrak{R}$ as a spanning subgraph;
\index{graph ! spanning}%
\item[(c)] $\mathscr{F}_\Gamma\subseteq\mathscr{F}_{\mathfrak{R}}$.
\end{itemize}
\end{proposition}

This result~\cite{tarzi1} shows that $\mathscr{F}_{\mathfrak{R}}$ is the unique maximal neighbourhood
filter.
\index{filter}%
 But this uniqueness is only up to isomorphism. So part (c) really means that $\mathscr{F}_\Gamma$ is contained in a filter isomorphic to $\mathscr{F}_{\mathfrak{R}}$.

\bigskip

Grafting QM onto classical objects such as random graphs
\index{graph ! random ($\mathfrak{R}$)}%
 can either be done by attaching Hilbert space(s) to vertices or edges, or as an integrand in a Feynman path integral.
\index{path integral}%
  Heuristicallly the random mathematical structure $\mathcal{M}$ can be interpreted as an affine space; its basis of vectors is itself an affine space which must be interpreted as the space of paths.  Each unit volume in the vector space has a weight-function-induced probability density for the infinitesimal region around a point to be realized as a true structure. 

\bigskip

The algebraic structures relevant to applications could be rings rather than vector spaces.  Rings are flexible enough to allow linear superpositions of elements, are more general than algebras, and intersect algebras in the class of linear spaces; see Appendix B of~\cite{tarzi1}.

Two relational structures
\index{structure ! relational}%
 with the same age have the same graded algebra up to isomorphism, hence the name \emph{age algebra}.
\index{age algebra}%
In fact the homogeneous components are indexed by an age of the appropriate size and so algebras built out of the age are equal and not just isomorphic.  The converse is not true.

In~\cite[Chapter 5]{tarzi1} we found an isomorphism between two algebras with very different structures in the finite algebra case, one graded and the other semisimple, but which are isomorphic in the infinite case.

Let $\mathcal{M}$ be a relational structure
\index{structure ! relational}%
 on a set $X$, finite or infinite. For any
natural number $n \in \mathbb{N}$, let $V_n$ denote the vector space of all complex-valued functions from $n$-element subsets of $X$ to $\mathbb{C}$ which are constant on isomorphism classes of $n$-element substructures of $\mathcal{M}$, and construct an algebra by
\[\mathcal{A}=\bigoplus_{n\ge0}V_n.\]

There are different ways to define multiplication on $\mathcal{A}$, each of which could turn it
into a commutative and associative algebra.  This was studied in~\cite{tarzi1}.

\smallskip

When the relational structure
\index{structure ! relational}%
 is a graph, there is a separate combinatorial  problem of how many graphs can be fitted into a larger graph on a given vertex set, and one approach to its resolution is exemplified by the following.  

Denoting the number of $i$-vertex graphs $\Gamma_i$ that can be embedded as induced subgraphs of a $j$-vertex graph ($j \geq i$) by a double angled bracket, we have that

\begin{equation}
\quad \llangle \Gamma_j : \Gamma_i \rrangle = \sum  \llangle \Gamma_j : \Gamma_{j - 1}  \rrangle \llangle \Gamma_{j-1} : \Gamma_i \rrangle
\end{equation}

\begin{equation}
= \left[ {j-1 \choose j-1} + \ldots + {j-1 \choose 2} + {j-1 \choose 1} \right] \llangle \Gamma_{j-1} : \Gamma_i \rrangle
\end{equation}

where the sum is over all $j-1$ vertex graphs.  This gives a recursive combinatorial expression, with the possible use of the following known identities:-

(1)  the recurrence relation for Bell numbers:\\
\index{Bell number}%
$\Bn_n = \sum_{k=1}^{n} {n-1 \choose k-1} \Bn_{n-k} = \sum_{k=1}^{n} S(n, k)$,
\index{Stirling number ! of second kind}%

(2)  ${n \choose k} = {n-1 \choose k-1} + {n-1 \choose k}$,

(3)  $L(n, k) = \left( \frac{n-1}{k-1} \right) \frac{n!}{k!}$;\quad $L(n, k+1) = \frac{n-k}{k(k+1)} L(n, k)$;\quad and\\ $(-1)^n L(n, k) = \sum_z (-1)^z s(n, k) S(n, k)$,

where the unsigned Lah number $L(n, k)$ counts the number of ways a set of $n$ elements can be partitioned into $k$ nonempty linearly ordered subsets, and $s(n, k)$ is the Stirling number of the first kind.
\index{Stirling number ! of first kind}%

\bigskip



\emph{Combinatorial Enumeration}

Let $\mathcal{M}$ be a countable homogeneous structure
\index{structure ! homogeneous}%
 with age
\index{structure ! age}%
 $\mathcal{A}$ and let $G = \Aut(\mathcal{M})$.  By homogeneity, two finite subsets of $\mathcal{M}$ lie in the same $G$-orbit if and only if the induced substructures are isomorphic.  So the sequence enumerating \emph{unlabelled} $n$-element members of $\mathcal{A}$ (that is, up to isomorphism) is identical with the sequence enumerating the $G$-orbits on unordered $n$-element subsets of $\mathcal{M}$.  Similarly, the number of  \emph{labelled} $n$-element members of $\mathcal{A}$ (that is, members of $\mathcal{A}$ on the set $\{1, 2, \ldots, n\}$) is equal to the number of $G$-orbits on ordered $n$-tuples of distinct elements of $\mathcal{M}$.

Let $G$ be an oligomorphic permutation group acting on each of the sets $X^n$ (all $n$-tuples of elements of $X$), $X^{(n)}$ (all $n$-tuples of distinct elements of $X$), and $X^{\{n\}}$ (all $n$-element subsets of $X$).  If $F_{n}^{*}, F_{n}$ and $f_n$ denote these numbers of orbits then $f_n \leq F_n \leq n! f_n$, since each orbit on $n$-sets corresponds to at least one and most $n!$ orbits on $n$-tuples.  A permutation group is \emph{$n$-transitive}
\index{group ! permutation ! $n$-transitive}%
 if $F_n = 1$,  and \emph{$n$-homogeneous}
\index{group ! permutation ! $n$-homogeneous}%
if $f_n = 1$.

Also,
\[ F_{n}^{*} = \sum^{n}_{k=1} S(n, k) F_{k}, \]
where $S(n, k)$ is the Stirling number of the second kind,
\index{Stirling number ! of second kind}%
 the number of partitions of an $n$-set into $k$ parts.  For an orbit $(\alpha_1, \ldots, \alpha_n)^G$ on $n$-tuples determines, and is determined by, a partition of $\{1, \ldots, n\}$ into $k$ parts (where $i$ and $j$ lie in the same part if $\alpha_i = \alpha_j$) and an orbit on $k$-tuples of distinct elements.  The \emph{exponential generating function} 
\index{exponential generating function}%
is given by $F(t) = \sum \frac{F_n t^n}{n!}$.  The series $F(t)$ for a direct product (in the intransitive action) or a wreath product (in the imprimitive action)
\index{group ! permutation ! imprimitive}%
 can be calculated from those of the factors:
\[ F_{G \times H} (t) =  F_{G} (t) \times F_{H} (t), \]
\[ F_{G \Wr H} (t) =  F_{H} (t) (F_{G} (t) - 1). \]

If $S = \Sym(\omega)$ where $\omega$ denotes countable infinity and the sequence $\alpha$ is realized by a group $G$ then $S \alpha$ is realized by $G \Wr S$.

Let $G$ act transitively on $X$, and $G_{x}$ be the stabilizer of the point $x \in X$.  It can be shown that 
\[ F_{G_{x}} (t) = {\operatorname{d}\over\operatorname{d}t} F_G (t), \]
or equivalently differentiating an exponential generating function corresponds to shifting the sequence terms one place to the left,
\[ F_{n} (G_x) = F_{n+1} (G). \]
So there is an equivalence between $G_x$-orbits on $n$-tuples and of $G$ orbits on $(n+1)$-tuples.  If $G$ is intransitive, then the derivative of $F_G (t)$ gives the sum of $F_{G_x} (t)$, over a set of representatives $x$ of the orbits of $G$.

For any oligomorphic group $G$, \emph{generalized Stirling numbers}
\index{Stirling number ! generalised}%
denoted by $S[G](n, k)$, can be defined that obey
\[ \sum^{n}_{k=1} S[G](n, k)  =  F_{n}(G \Wr S),\]
and have the composition property
\[ \sum^{n}_{l=k} S[G](n, l) S[H](l, k)  =  S[G  \Wr H](n, k),\]
where $\Wr$ denotes wreath product.  The last result can be expressed in terms of infinite lower triangular matrices of generalized Stirling numbers.  

There is a linear analogue of the combinatorics
\index{combinatorics}%
 of sets and functions that applies to vector spaces over finite fields and linear transformations in which the Gaussian (or $q$-binomial coefficient) replaces the binomial coefficient.  These coefficients enumerate the number of $k$-dimensional subspaces of $n$-dimensional $GF(q)$-vector space~\cite{kacc}.

\bigskip

There are general theorems on growth rates of counting sequences, many of them requiring a primitive permutation group
\index{group ! permutation ! primitive}%
 $G = \Aut(\mathcal{M})$, for example if $\mathcal{M} = \mathfrak{R}$.   Primitivity can be completely described if $G$ is oligomorphic.

\bigskip

Within the area of probabilistic combinatorics, there are various concepts with suggestive titles such as the \emph{concentration of measure} in discrete random processes that may be brought to bear but these would have to be studied before an assessment can be made of their utility to Random Dynamics
\index{Random Dynamics}%
 program.

With one eye on the philosophy of the path integral,
\index{path integral}%
 one of the questions in random graph
\index{graph ! random ($\mathfrak{R}$)}%
 theory is can it be shown that some random variables are concentrated around either their expected values or their expected trajectories as an underlying random process evolves?  The differential equation method of Wormald~\cite{wormald} 
\index{Wormald, N.}%
  and its extension due to Bohman~\cite{bohman}
\index{Bohman, T.}%
uses solutions of differential equations to approximate the dynamical evolution of the random process.

Once space and time are derived position, momentum and translational invariance follow as well as energy and momentum conservation.  Rotational
\index{group ! orthogonal (rotation)}%
 invariance may come with angular momentum conservation.

\bigskip

Nielsen and Kleppe
\index{Nielsen, H. B.}%
\index{Kleppe, A.}%
 returned to the theme of a derivation of space in~\cite{nielsenkleppe}.  They postulate the existence of an abstract, general phase space or state space together with a random generic Hamiltonian $H$ and then examine the statistically expected functional form of the ``random $H(\overrightarrow{q}, \overrightarrow{p})$''.  Within phase space one state is identified as the key vacuum state
\index{vacuum}%
 given by a wave function; classically this is a point but quantum mechanically it is a volume $h^{N}$ (uncertainty principle).
\index{uncertainty principle}%
  Because of quantum mechanics, space is identified with half of the $2N$ dimensions of the phase space of a very extended wave packet which is a superposition of the different field degrees of freedom. 

A wave function can be approximated by excitable displacements of the transversal directions of the $N$-dimensional manifold. 

The primacy of 3+1 dimensions is hinted at by the observation that if non-Lorentz
\index{Lorentz invariance}%
 invariant terms are added to the Weyl equation,
\index{Weyl equation}%
 only in 3+1 dimensions does Lorentz invariance eventually emerge.  
\index{Lorentz invariance}%

\smallskip

Random Dynamics
\index{Random Dynamics}%
 takes nature to be nonlocal because it is argued that locality only makes sense in the presence of a spacetime or at least a manifold, whilst field theory assumes locality that is all interactions occur in one spacetime point. 

This nonlocality is intended to be inherent in nature, as opposed to being of a dynamical origin, for example the purported quantum non-separability which arises as nonlocal correlations.  Locality is then explained using diffeomorphism symmetry (that is invariance under reparametrization mappings) and a Taylor expansion of the action.  Translational invariance also follows from diffeomorphism symmetry.
\index{diffeomorphism}%

\chapter{The Gauge Glass}
\index{gauge ! glass}%
Recall the postulate of the Random Dynamics
\index{Random Dynamics}%
 project that any sufficiently general initial assumptions or axioms or model at the fundamental scale will lead to the same known physical laws.

Nielsen
\index{Nielsen, H. B.}%
 and co-workers have devised discrete spacetime so-called \emph{gauge glass}
\index{gauge ! glass}%
 models, inspired by lattice gauge theory,
\index{lattice gauge theory (LGT)}%
 to derive criteria for the breaking of a large over-group to the Standard Model
\index{Standard Model (SM)}%
 group.  The name \emph{gauge glass} was chosen to be analogous to a spin glass whose ``frozen in'' structure is reminiscent of that of glass.

An explanation as to why the SM group is 
\index{Standard Model (SM)}%
  group $\S(\U(2) \times \U(3))$ $(\subseteq \SU(5))$
\index{group ! SU(5)@$\SU(5)$}%
\index{group ! S(U(2) \times U(3))@$\S(\U(2) \times \U(3))$}%
defined by the set of matrices and why groups with complex representations are likely to be more important down to low energies than those with only real self-conjugate representations can be found in the paper~\cite{bennettbrene} by Bennett, Brene and Nielsen.  
\index{Bennett, D. L.}%
\index{Brene, N.}%
\index{Nielsen, H. B.}%
 The complex representations should have complex conjugation as an 
\index{automorphism}%
automorphism; it is the only automorphism of the SM group.

A reflexive, symmetric and transitive binary relation on a set is an equivalence relation, and one that is reflexive, antisymmetric and transitive is a partial order.   A \emph{lattice}
\index{lattice}%
 in \emph{lattice gauge theory}
\index{lattice gauge theory (LGT)}%
 is a set of elements with a partial order, the elements being lattice \emph{sites} connected by lattice \emph{links}.  The sites can be interpreted as charges or as sinks or sources of magnetic fields.   

Link variables defined on the edges of the lattice
\index{lattice}%
 are the fundamental variables of the lattice theory.  Denoted $U(\xymatrix{{\bullet}_x \ar@{-}[r] & {\bullet}_y)} \in G$ they are elements of the gauge group $G$ describing a symmetry of the lattice gauge theory.  The continuum space version of the link variable is a parallel transport operator between the points $x$ and $y$:
\[ U(x, y) = \text{P} e^{i g \int_{C} A_{\mu} (x) dx^{\mu}} \]
 where \text{P} is the path ordering operator and $C$ is a curve between points $x$ and $y$.  The operator $W = tr( \text{P} e^{i g \oint_{C} A_{\mu} (x) dx^{\mu}})$ is the Wilson loop.  For a scalar field $\phi(x)$ interacting with a gauge field $A_{\mu}$ the term $\phi^{+}(y) U(x, y) \phi(x)$ is a gauge invariant observable.

The link variable connects the point $n$ and the point $n + a_{\mu}$, where the index $\mu$ indicates the direction of a link in the hypercubic lattice
\index{lattice}%
 with parameter $a$.  The lattice version of the equation for the parallel transport operator is
\[ U(\xymatrix{{\bullet}_x \ar@{-}[r] & {\bullet}_y)} = e^{i \theta_{\mu} (n)} \equiv U_{\mu} (n). \]
Considering the continuum limit $a \to 0$ of this operator we have that $\theta_{\mu} (n) = a \hat{A}_{\mu}(x)$ where $\hat{A}_{\mu}(x) = g A^{j}_{\mu} (x) t^j$ and $t^j$ generates $G = \SU(n)$.  For example, if $n =1$ then $\hat{A}_{\mu}(x) = g A_{\mu}(x)$ and for $n=3$ $t^j = \lambda^j / 2$ where $\lambda^j$ are the Gell-Mann matrices.

\begin{figure}[!h]$$\xymatrix{
&& {}  \ar@{-}[d] && {} \ar@{-}[d] \\
& {} \ar@{-}[r]  & {\bullet}  \ar@{-}[dd]_{l_4} \ar@{-}[rr]^{l_3} && {\bullet} & {} \ar@{-}[l] \\
\\
& {} \ar@{-}[r]  & {\bullet} \ar@{-}[rr]_{l_1} && {\bullet} \ar@{-}[uu]_{l_2}  & {} \ar@{-}[l] \\
&& {}  \ar@{-}[u] && {} \ar@{-}[u]
}$$
\caption{Lattice Gauge Theory Plaquette}
\label{latgth}
\end{figure}
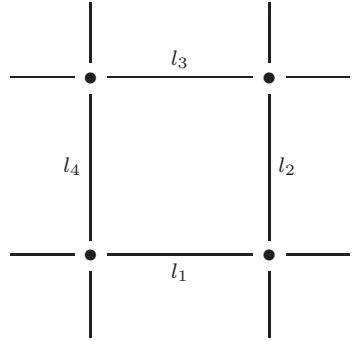

The links couple together to form lattice
\index{lattice}%
 \emph{plaquettes}, denoted $\square$, where the plaquette variables $U_{\square}$ and plaquette action $\mathcal{S}_{\square}$ contributions are defined below; see Figure~\ref{latgth}.  To model the gauge invariant
\index{gauge ! invariance}%
 action of the gauge theory assume that Bianchi identities are satisfied; see next chapter.  These identities apply also to plaquettes in more than 2 dimensions, for example to cubes in a $3$-dimensional lattice.
\index{lattice}%

The lattice action $\mathcal{S}$ is invariant under the gauge transformations
\[ U(\xymatrix{{\bullet}_x \ar@{-}[r] & {\bullet}_y)} \to \Lambda(x) U(\xymatrix{{\bullet}_x \ar@{-}[r] & {\bullet}_y)} \Lambda^{-1}(y) \]
on a lattice where $\Lambda(x) \in G$.  In 4-dimensional Euclidean space the analogue of the partition function is a path integral
\[ Z = \int DU(\xymatrix{{\bullet} \ar@{-}[r] & {\bullet}}) e^{-S[U(\xymatrix{{\bullet}_x \ar@{-}[r] & {\bullet}_y)}]}. \]

The lattice field theory is built in such a way that in the continuum limit it gives a regularized smooth gauge theory of fields $A^{j}_{\mu} (x)$.

The coupling constant strength determines the quantum field fluctuations which in turn determine the vacuum
\index{vacuum}%
 phases.  Short range correlations correspond to the confinement-like phase in which the Bianchi identities can be assumed negligible implying the approximate independence of the fluctuations of each plaquette in a lattice cell.
\index{lattice}%
  In the Coulomb phase the electromagnetic potential and photons have infinite range and so the Bianchi identities lead to a correlation of plaquette variable fluctuations over distances of many lattice constants and more.

Different combinations of gauge field 
\index{gauge ! field}%
phases correspond to different vacuua.  The phase transations at the \emph{multiple point} of a phase diagram for a lattice gauge theory
\index{lattice gauge theory (LGT)}%
 are first order and the phases are distinguishable at the lattice scale.
\index{lattice}%

Commutator relations in non-abelian groups
\index{group ! non-abelian}%
 constrain the gauge coupling normalization but they are absent in abelian
\index{group ! abelian}%
 groups.  Rescalings of the gauge
\index{gauge}%
 potential are only possible in the abelian sectors so here the gauge couplings have no physical significance and so the $\U(1)$ has no intrinsic unit of charge.

By analogy with lattice gauge theory,
\index{lattice gauge theory (LGT)}%
 consider links between sites and associate group-valued link variables corresponding to a gauge group
\index{group ! gauge}%
 $G$ which gives symmetries of the theory.  Introduce dynamics through an action $\mathcal{S}$ composed additively of contributions from all primitive polygons (plaquettes in the case of a regular cubic lattice).  
\index{lattice}%
To achieve gauge invariance
\index{gauge ! invariance}%
 on the lattice, the individual ``plaquette'' contributions $\mathcal{S}(\square)$ must be invariant functions of the ``plaquette'' elements which themselves are products of the link elements circumscribing the ``plaquettes'':
\begin{equation}
U_{\square} : = U(\square_1) U(\square_2) U(\square_3) U(\square_4),
\end{equation}

\[ \mathcal{S}_{\square} = \mathcal{S}(class (U(\square))). \]
The class of an element here refers to the conjugacy class set of all elements equivalent to the element under conjugation:
\[class(U) = \{g U g^{-1} : g \in G\}.\]

An example of a class function is $\mathcal{S}_{\square} = Re [ \beta_{\square} tr_{rep} \{ \U(\square) \} ]$ where $tr_{rep}$ denotes the trace of some matrix representation, and $\beta = 1 / g_c^2$ where $g_c$ is the gauge
\index{gauge}%
 coupling constant.

In the simplest case, $G = U(1)$, $\mathcal{S}_{\square} = \beta \sum_{\square} Re ( \U(\square) )$ and the link variables $U(\xymatrix{{\bullet} \ar@{-}[r] & {\bullet})} \in \mathbb{C}$ are modulus one complex numbers.

In the lattice model, the Lorentz gauge condition is 
\[ \prod U(\xymatrix{{\bullet}_x \ar@{-}[r] & {\bullet}_y)} = 1 \] 
where the product is over all links $\xymatrix{{\bullet}_x \ar@{-}[r] & {\bullet}_y}$.  Also we can write $U(\xymatrix{{\bullet}_x \ar@{-}[r] & {\bullet}_y)} = e^{i \theta_{\square}}$.  These plaquette link variables are not independent but satisfy a Bianchi identity (see later).  So the simplest lattice $U(1)$ action takes the form $\mathcal{S}_{\square} = \beta \sum_{\square}  cos(\theta_{\square})$.  For the compact lattice QED, $\beta = 1 / e^{2}_0$ where $e_0$ is the bare electric charge.

Wilson's original simplest action for the lattice $SU(n)$ gauge theories $\mathcal{S} = - \frac{\beta}{n} \sum_{\square} Re(tr(U_{\square}))$ where the sums run over all plaquettes of a hypercubic lattice and $U_{\square}$ belongs to the adjoint representation of $SU(n)$ generalized by Bhanot and Creutz~\cite{bhanot1}~\cite{bhanot2}
\index{Bhanot, G.}%
\index{Creutz, M.}%
 to
\[ \mathcal{S} = \sum_{\square} [ - \frac{\beta_f}{n}  Re(tr(U_{\square}))  - \frac{\beta_A}{N^2 - 1}  Re( tr_A(U_{\square})) ] \]\\
where $\beta_f, tr$ and $\beta_A, tr_A$ are respectively the lattice constants and traces in the fundamental and adjoint representations of $SU(n)$.

The phase diagrams for the generalized lattice $SU(2)$ and $SU(3)$ theories using Monte Carlo methods have a triple (boundary) point of three first-order phase transitions, where the ``Coulomb-like'', $Z_n$ and $SU(n) / Z_n$ confinement phases meet.  Three phase border lines emanate from the triple point which separate the corresponding phases.  The $SU(n) / Z_n$ phase transition is due to a condensation of monopoles.  The discrete $Z_n$ phase transition occurs when lattice plaquettes change from the identity to nearby elements in the group.

Lattice gauge theory
\index{lattice gauge theory (LGT)}%
 assumes a regular lattice, a given gauge group
\index{group ! gauge}%
 and a given form of the plaquette action throughout the lattice.  A gauge glass
\index{gauge ! glass}%
 model has randomly varying parameters on the ``lattice'', though the same random value at any given plaquette is retained throughout, that is it is ``quenched''.

Randomness can be introduced in a variety of ways:-

(i) The sites can be irregularly distributed (hence the quotation marks around ``lattice'' and ``quenched'').  However this possibility is not considered and the gauge glass
\index{gauge ! glass}%
 lattice
\index{lattice}%
 is taken to be a regular hypercubic lattice in $n$-dimensional space (-time).

(ii)  Fix the gauge group
\index{group ! gauge}%
 throughout the lattice but randomly vary the functional form of each plaquette action, for example in 
\[ \mathcal{S} = \sum_{\square} Re [ \beta_{\square} tr_{q} \{ \U(\square) \} ] \]take $\beta_{\square}$ to have a random value for each plaquette and the trace to be in the quark-like representation $q$.

(iii)  If the gauge group
\index{group ! gauge}%
 has isomorphic subgroups, which are present on all links that join at a site, these are not in the first instance distinguishable in a gauge glass
\index{gauge ! glass}%
 model.  So in going around a plaquette the subgroups can either remain separate or can be mixed together.

(iv)  Associate a new group in a random quenched manner to each link; again this is not considered in the gauge glass models.
\index{gauge ! glass}%

\smallskip

Regardless of the randomness, the action is constructed to be invariant under the gauge group
\index{group ! gauge}%
 $G$ but the ground vacuum
\index{vacuum}%
 state, which is determined by the minimum energy value, need not be.  Let $\U(\square)_{pref}$ be the energetically preferred value of plaquette group element.  Only if $\U(\square)_{pref}$ commutes with every element of the group $G$ will the vacuum be invariant under global gauge transformations.
\index{gauge ! transformation}%
  So the best surviving elements will belong to the center of the group,
\index{group ! center}%
 that is the set of elements that commute with each group element.  The $\SU(n)$
\index{group ! SU(n)@$\SU(n)$}%
 groups have such central elements, but the $\SO(n)$
\index{group ! SO(vn)@$\SO(n)$}%
 groups have a discrete part of the center divided out.  The center of all groups of the form $\SO(2n)$ $n > 1$, for example $\SO(8)$
\index{group ! SO(8)@$\SO(8)$}%
 is $\mathbb{Z}_2 = \{ \pm 1\}$.

The central elements are required to be continuous meaning that there are elements in the center
\index{group ! center}%
 that are arbitrarily close to the unit element.  Also the center must be connected so that the preferred plaquette values do not lie in a different nonconnected part of the center.  The group $\S(\U(2) \times \U(3))$
\index{group ! S(U(2) \times U(3))@$\S(\U(2) \times \U(3))$}%
 has a compact connected center.

The purported randomly chosen gauge group
\index{group ! gauge}%
 near the Planck energy breaks down in a series of steps until the group remaining at the gauge glass
\index{gauge ! glass}%
 level has the required properties of its central elements and has no multiple occurrences of isomorphic factors.

According to the Random Dynamics philosophy,
\index{Random Dynamics}%
 in reaching the group $\S(\U(2) \times \U(3))$
\index{group ! S(U(2) \times U(3))@$\S(\U(2) \times \U(3))$}%
 neither the specific structure of the ``lattice''
\index{lattice}%
 nor the choice of action for any given plaquette nor the group chosen as long as it is sufficiently large, are crucial.

\medskip

As to the origin of the gauge glass,
\index{gauge ! glass}%
\index{Foerster, D.}%
\index{Nielsen, H. B.}%
\index{Ninomiya, M.}%
 associated to each link of a random lattice
\index{lattice}%
 group variables which approximately stabilize the local action contributions~\cite{foerster}.  These new variables may eventually lead to gauge fields
\index{gauge ! field}%
 with massless gauge
\index{gauge}%
 quanta for those parts of the symmetry that are not broken.  In other words the large distance behaviour of gauge theories
\index{gauge ! theory}%
 is stable, within certain limits, to the addition of gauge non-invariant interactions at small distances.

\medskip


\emph{Switchings and Switching Automorphisms of Graphs}
\index{automorphism ! switching}%

Having proved theorems~\cite{tarzi1}~\cite{tarzi2} that connect random graph vertices and edges in a $1$--$1$ correspondence with lattice
\index{lattice}%
 points and links respectively we require at least one physical interpretation of the graph vertices and edges.  We are potentially provided with one by the Random Dynamics
\index{Random Dynamics}%
 program where the regular hypercubic lattice in normal lattice gauge theory is replaced by an amorphous glass-like structure simply by letting the lattice sites take random positions separated by distances of order the Planck length.  (It could be that a lattice is the local version of a global random graph structure, in parallel to the Random Dynamics
\index{Random Dynamics}%
 philosophy that what we observe at our low energy level can be interpreted as essentially correction terms of a description of the physics taking place at a higher energy level.)  This interpretation needs to be expanded upon, but one possible step thereafter is to understand some of the groups that are naturally supported by random graphs and then to explore whether or not any physical meaning can be attributed to such groups. The purpose of this subsection is to look at one type of such groups.

\smallskip

Let $\mathcal{G}_{m,n}$ be the set of simple complete graphs
\index{graph ! complete}%
 on a fixed set of $n$ vertices and $m$ edge-colours and $\Gamma \in \mathcal{G}_{m,n}$ be an arbitrary element of this set.

Let $c$ and $d$ be distinct colours, and $Y$ a subset of the vertex
set~$X$ of the graphs in $\mathcal{G}_{m,n}$. The operation
$\sigma_{c,d,Y}$ interchanges the colours $c$ and $d$ whenever
they occur on an edge with just one vertex in $Y$, leaving all other colours
unchanged. Note that
\[\sigma_{c,d,Y} = \sigma_{d,c,Y} = \sigma_{c,d,X\setminus Y}\]
and
\[\sigma_{c,d,Y}^2=1.\]
The \emph{switching group}
\index{group ! switching}%
 $S_{m,n}$ is the group of permutations of $\mathcal{G}_{m,n}$ generated by the \emph{switching operations}, which are defined to be the switchings on all vertex subsets of $X$. A \emph{switching class} is an orbit of $S_{m,n}$ on $\mathcal{G}_{m,n}$.

The structure of the group of all switchings on $n$-vertex subgraphs having $m$ colours
is given by
\[S_{m,n} \cong (\Alt(m))^{n(n-1)/2}\sd (C_2)^{n-1}.\]
Here $C_2$ is the smallest non-trivial group consisting of the identity element and transpositions, and the alternating group
\index{group ! alternating ($\Alt$)}%
 $\Alt(m)$ of even permutations is the index-2 normal (that is, invariant) subgroup
\index{group ! normal (invariant) subgroup}%
 of the symmetric group $\Sym(m)$
\index{group ! symmetric ($\Sym$)}%
 of all permutations.

The semi-direct product
\index{group ! semi-direct product}%
 action here partitions the vertex set into two parts and the $C_2$ groups transpose the colours on edges crossing the partition.  This structure is retained in the limit of an infinite number of vertices, as $n \to \infty$, and it specializes to $S_{2,n} = C_2^{n-1}$ in the two-colour case.

If $g$ is any element of $\Sym(X)$, the symmetric group
\index{group ! symmetric ($\Sym$)}%
 of all permutations on $X$, then $\sigma_{c,d,Y}^g=\sigma_{c,d,Y^g}$.
So $\Sym(X)$ normalises $S_{m,n}$.
\index{group ! normal (invariant) subgroup}%
 Note that $\Sym(X)$ also acts on the set $\mathcal{G}_{m,n}$, and hence so
does the semi-direct product of these groups.
\index{group ! semi-direct product}%

Let $\Gamma$ be an element of $\mathcal{G}_{m,n}$, that is, an $m$-coloured
graph on $X$. The \emph{group of switching automorphisms}
\index{group ! switching automorphism}%
 $\SAut(\Gamma)$ is defined by
\[\SAut(\Gamma) = \{g\in\Sym(X) : \exists \sigma\in S_{m,n},\ \Gamma g=\Gamma\sigma\}. \]
That is, it consists of all permutations of $X$ whose effect on $\Gamma$ can
be undone by a switching.

Another way to think about this group is as follows: if $\hat{G}$ is the
stabilizer
\index{group ! stabilizer}%
 of $\Gamma$ (so $\Gamma = \Gamma \sigma g^{-1}$, and $g$ undoes
the effect on $\Gamma$ of $\sigma$) in the semi-direct product
\index{group ! semi-direct product}%
$S_{m,n} \rtimes \Sym(X)$, then $\SAut(\Gamma)$ is the image
of $\hat{G}$ under the canonical projection of the semi-direct product
\index{group ! semi-direct product}%
onto $\Sym(X)$.

Note that multicoloured graphs
\index{graph ! random ! $m$-coloured}%
 in the same switching class have the same group of
switching automorphisms.
\index{automorphism ! switching}%
 For suppose that $g\in \SAut(\Gamma)$, so that
$\Gamma g = \Gamma\sigma$, and let $\tau$ be any switching operation. Then
$(\Gamma\tau)g = (\Gamma\tau) \tau\sigma\tau^g$; since $\tau^g$ is again a
switching operation, $g$ is a switching automorphism
\index{automorphism ! switching}%
 of $\Gamma\tau$.

That switching is an equivalence relation is easily seen: (a)
Reflexivity - switch with respect to the whole of set $X$ and
everything outside $X$, that is $\emptyset$. (b) Symmetry - switch
with respect to $Y \subset X$; and then $Y$ again to recover the
original configuration. (c) Transitivity - switch with respect to $Y_1
\subset X$ then with respect to $Y_2 \subset X$ is the same as
switching with respect to $Y_1 \triangle Y_2$, the symmetric difference of $Y_1$ and $Y_2$.

\emph{A First Presentation of $S_{m,n}$}.
In~\cite[Chapter~4]{tarzi1}, we derive two presentations of the switching groups
\index{group ! switching}%
 $S_{m,n}$ for graphs on $n$ vertices and $m$ colours. 
The first comes out as

$S_{m,n} = \langle \sigma_{c,d,i}, 1 \le i \le n, 1 \le c < d \le m,\ |\
\sigma_{c,d,i}^2 = 1, \forall c, d, i,\\ (\sigma_{c,d,i}
\sigma_{c,d,j})^2=1, \forall c, d, (i \ne j),\ (\sigma_{c,d,i}
\sigma_{c',d',i})^3=1, \forall i\ \forall \{c,d\} \ne \{c',d'\},\\
(\sigma_{c,d,i} \sigma_{d,c',j})^2=(\sigma_{c,c',i} \sigma_{c,d,j})^2
\ ( = (\sigma_{c,d,i} \sigma_{d,c',j} \sigma_{c,d,i}
\sigma_{c,d,j})^2),\ (\forall c \ne d \ne c')\\ (\forall i<j) \rangle$.

\bigskip

The second presentation leads us to define and study a type of generalized Coxeter group, with the following as basic motivation.
\index{group ! Coxeter}%

A \emph{Coxeter group} 
\index{group ! Coxeter}%
has generators $s_1, \ldots, s_k$ and relations $s_i^2 = 1$ and $(s_i s_j)^{m_{ij}} = 1$, where $m_{ij} = \{2, 3, \ldots, \infty\}$.  The finite Coxeter groups are the finite real reflection groups~\cite{hum}.  
\index{group ! reflection}%

The Coxeter group $\Cox(\Gamma)$ may be defined on a graph $\Gamma$ and is generated by vertices $s_i \in \Gamma$ $(i \in I)$, such that
\begin{displaymath}
\begin{array}{ll}
s_i^{2}=1 & \textrm{$\forall s_i \in \Gamma$};\\
(s_i s_j)^2 = 1 \Leftrightarrow s_i s_j = s_j s_i & \textrm{if $s_i, s_j$ are disjoint};\\
(s_i s_j)^3 = 1 \Leftrightarrow s_i s_j s_i = s_j s_i s_j & \textrm{if $s_i, s_j$ are joined by a single edge};\\
(s_i s_j)^{m_{ij}} = 1& \textrm{where an edge is labelled $m_{ij}$ or}\\
{} & \textrm{there are $m_{ij} - 2$ parallel edges.}
\end{array}
\end{displaymath}\\
For example, $\Sym(n)$ is a Coxeter group represented by the following diagram:
$$\xymatrix{
{\circ} \ar@{-}[r] & {\circ} \ar@{-}[r] & {\circ} \ar@{-}[r] & {} {\ldots} {\ldots} \ar@{-}[r] & {\circ}
}$$
which has $n-1$ nodes indexed by $(i \in I)$, and where $s_i \mapsto (i, i+1)$.

\bigskip

\emph{A Second Presentation of $S_{m,n}$}.  A second type of $S_{m,n}$ presentation is given in terms of
involutions about vertices and $3$-cycles about edges, and can be recovered
by inspection of the form of the switching groups.
\index{group ! switching}%
  For example $S_{3,3} \cong
(\Alt(3))^{3}\sd (C_2)^{2}$ consists of $3$ generators $a_1, a_2, a_3$ that are intended to
represent three $3$-cycles and two involutions $b_1$ and $b_2$ whose
actions are shown in the following picture
$$
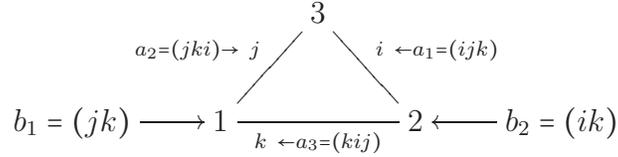
\xymatrix{ 
&& {3} \ar@{-}[dl]_{a_2=(j k i) \rightarrow \ j}  \ar@{-}[dr]^{i\ \leftarrow a_1=(i j k)}\\
{b_1=(jk)} \ar@{->}[r] & {1} \ar@{-}[rr]_{k\ \leftarrow a_3 = (k i j)} && {2} & {b_2=(ik)} \ar@{->}[l]
}$$
The presentation is given by

$S_{3,3} = \langle a_1, a_2, a_3, b_1, b_2\ |\ a_i^3 = 1,\ [a_i, a_j] =
1,\ b_i^2 = 1,\ [b_i, b_j] = 1,\\ b_1^{-1} a_1 b_1 = a_1,\ b_1^{-1} a_2 b_1
= a_2^{-1},\ b_1^{-1} a_3 b_1 = a_3^{-1},\ b_2^{-1} a_1 b_2 = a_1^{-1},\
b_2^{-1} a_2 b_2\\ = a_2,\ b_2^{-1} a_3 b_2 = a_3^{-1} \rangle$.

Extrapolating from $3$ to a finite number $n$ of vertices, we can give
generators for $S_{3,n}$ as $b_1, \ldots, b_{n-1}, a_{ij}, 1 \leq i <
j \leq n$, where the $b_i$ are transpositions of $2$ colours on edges
$i, j$ for all $j \ne i$ and the $a_{ij}$ are $3$-cycles on colours on
edge $\{i, j\}$.  The relations in the presentation are:

$a_{ij}^{3}=1$,\quad $[a_{ij},a_{kl}]=1$ $(i,j \neq k, l)$,\quad $b_i^{2}=1$,\quad $[b_i,b_{j}]=1$ $(i \ne j)$,
\begin{displaymath}
b_i^{-1} a_{jk} b_i = \left\{ \begin{array}{ll}
a_{jk}^{-1} & \textrm{if $j=i$ or $k=i$}\\
a_{jk} & \textrm{if $i \notin \{j,k\}$.}
\end{array} \right.
\end{displaymath}

Certainly $S_{3,n}$ satisfies this presentation.  But we must show
that this gives a defining set of relations for $S_{3,n}$.  If
$G$ is a group defined by this presentation then $|G| \leq |S_{3,n}|$
and we also have the surjection $G \twoheadrightarrow S_{3,n}$.  Therefore we have equality and a verification that this is a presentation for $S_{3,n}$.

\bigskip

We can extrapolate further from $m=3$ to any finite number $m$ of colours, and
use the presentation

\centerline{$\Alt(m)= \langle a_1, \ldots, a_{m-2} |  a_k^{3}=1,\ (a_k a_l)^2=1\ (k \neq l) \rangle$.}
\index{group ! alternating ($\Alt$)}%

This gives the generators for $S_{m,n}$ as: $b_i$ where $b_i$ is a $2$-cycle $(m-1, m)$ on colours on all edges $\{i, j\}$ for all $1 \le i < j \le n$, and $a_{ij}^{k}$ for $1
\le k \le m-2$ which are $3$-cycles $(k, m-1, m)$ of colours on all edges $\{i, j\}$.  Then
$(a_{ij}^{k} a_{ij}^{l})^2 = ( (k, m-1, m)(l, m-1, m) )^2
= ( (k\ m)(l\ m-1) )^2 = 1$.  The relations are:

$(a_{ij}^{k})^{3}=1,\quad (a_{ij}^{k}\ a_{ij}^{l})^{2} = 1\ (l \neq k),\quad [a_{ij}^{k},\ a_{i'j'}^{l}] = 1\ (\text{if}\ \{i',j'\} \neq \{i,j\}),\quad b_i^{2}=1,\ [b_i, b_{i'}] = 1\ (\text{if}\ i \ne i')$,
\begin{displaymath}
b_h^{-1} a_{ij}^{k} b_h = \left\{ \begin{array}{ll}
(a_{ij}^{k})^{-1} & \textrm{if $h=i$ or $h=j$}\\
a_{ij}^{k} & \textrm{if $h \notin \{i,j\}$.}
 \end{array} \right.
\end{displaymath}

Each edge $\{i, j\}$ represents one alternating group
\index{group ! alternating ($\Alt$)}%
 in the direct product expansion of $S_{m,n}$.  To prove that this presentation defines precisely $S_{m,n}$, first recall the presentation 

$\Sym(m) = \langle a_{ij}^1, \ldots, a_{ij}^{m-2},\ b_{i} : (a_{ij}^k)^3 = 1 = (a_{ij}^k a_{ij}^l)^2\ (i \ne j),\\ b_{i}^2 = 1,\ b_{i} a_{ij}^k b_{i} = (a_{ij}^k)^{-1} \rangle$ for fixed $i, j$ and for all
$k$.  

Making the substitutions $a_{ij}^{k} \mapsto (k\ m-1\
m),\ b_{i} \mapsto (m-1\ m)$ we see that the $a_{ij}^{k}$ generate $\Alt(m)$. Then
proceeding as for $S_{3,n}$ proves the presentation.

\bigskip

How might this impact applications?  The obvious question now is that although a large variety of groups can be supported by the graph edges and vertices, the generators $a_{ij}$ and $b_i$ of the switching groups
\index{group ! switching}%
 and the groups themselves arise naturally as \emph{reducts} (or closed over-group of the automorphism group)
\index{group ! automorphism}%
 and given the $1$--$1$ correspondence between graph vertices and edges and lattice sites and links, do these generators or switching groups have a natural application in lattice gauge theory or gauge glass theory?

\bigskip

\emph{Formal appearance of gauge symmetry by definition}

Regardless of whether or not gauge symmetry is a true symmetry, it has been a central notion in the formulation of the field theories that have had great success in predicting and explaining a good deal of fundamental physics.  So we give a summary of the purely formal derivation of gauge symmetry appears in~\cite[p.~110]{froggatt}.

Define a Euclidean action for $\U(1)$
\index{group ! U(1)@$\U(1)$}%
 lattice
\index{lattice}%
 electrodynamics
\begin{equation}
\mathcal{S} = \beta \sum_{\square} Re(U_{\square}) + \alpha \sum_{-} Re(U(-)),
\end{equation}
 on a regular hypercubic Euclidean
\index{Euclidean space}%
 space-time lattice with lattice spacing $a$.  Define on each link, denoted $-$, the variables $U(-) \in U(1) = \{z : z \in \mathbb{C}, |z|=1 \}.$

The flux variables $U_{\square}$, for the plaquette denoted $\square$, is given by the product of four link variables on neighbouring links. If $\alpha = 0$ then with $\beta = \frac{1}{g_c^2}$ where $g_c$ is the gauge
\index{gauge}%
 coupling constant, the first term of equation (1) gives a pure $U(1)$ gauge theory
\index{gauge ! theory}%
 which is invariant under the lattice gauge transformation $\Lambda$, through
\index{gauge ! transformation}%
\begin{equation}
U(\xymatrix{
{\bullet}_x \ar@{-}[r] & {\bullet}_y) \to \Lambda(x)} U(\xymatrix{{\bullet}_x \ar@{-}[r] & {\bullet}_y) \Lambda^{-1}(y)}
\end{equation}

Here $y = x + a \delta_{\mu}$ and $\mu$ refers to the direction of the link $$\xymatrix{ {\bullet}_x \ar@{-}[r] & {\bullet}_y }$$ connecting the sites with coordinates $x^{\rho}$ and $x^{\rho} + a \delta_{\mu}^{\rho}$.

Gauge invariance
\index{gauge ! invariance}%
 is formally introduced by hand (subscript `h') by expressing the action in terms of field variables 
\[ U_h (\xymatrix{ {\bullet}_x \ar@{-}[r] & {\bullet}_y) \in \U(1) },\quad  H(\xymatrix{  \bullet}) \in U(1) \]
defined respectively on the links and sites, and where
\begin{equation}
U(\xymatrix{
{\bullet}_x \ar@{-}[r] & {\bullet}_y) : = H^{-1}(x)} U_h (\xymatrix{{\bullet}_x \ar@{-}[r] & {\bullet}_y) H(y).}
\end{equation}
The link variables are functions of oriented variables:
\begin{equation}
U_h(\xymatrix{
{\bullet}_x \ar@{<-}[r] & {\bullet}_y)  = } U_h^{-1} (\xymatrix{{\bullet}_x \ar@{->}[r] & {\bullet}_y) = } U_h^{*} (\xymatrix{{\bullet}_x \ar@{->}[r] & {\bullet}_y)
.}
\end{equation}
The new site variable $H(\xymatrix{  \bullet}) $ could represent a Higgs field.

The original field $U (\xymatrix{ {\bullet}_x \ar@{-}[r] & {\bullet}_y })$ and action $\mathcal{S}[ U (\xymatrix{ {\bullet}_x \ar@{-}[r] & {\bullet}_y })]$ which is a function only of this field are both invariant under the artificial gauge
\index{gauge ! transformation}%
 transformation $\Lambda(x) \in U(1)$:
\begin{equation}
U_h (\xymatrix{
{\bullet}_x \ar@{-}[r] & {\bullet}_y) \to \Lambda^{-1}(x)} U_h (\xymatrix{{\bullet}_x \ar@{-}[r] & {\bullet}_y) \Lambda(y),}\
 H(x) \to \Lambda^{-1} (x) H(x).
\end{equation}

Using the invariance of $U_{h\square} = U_{\square}$ of equation (3) under a $\U(1)$ gauge
\index{gauge ! transformation}%
 transformation, the action in equation (5) for $\U(1)$
\index{group ! U(1)@$\U(1)$}%
 lattice
\index{lattice}%
 electrodynamics exhibits formal gauge invariance
\index{gauge ! invariance}%
 in terms of the new variables as:
\begin{equation}
\mathcal{S}[U_h, H] = \beta \sum_{\square} Re[U_{h \square}] + \alpha \sum_{-} Re[H^{-1}(x) U_h(\xymatrix{ {\bullet}_x \ar@{-}[r] & {\bullet}_y) H(y)}].
\end{equation}

\smallskip

The \emph{mean field approximation (MFA)}
\index{mean field approximation (MFA)}%
 of lattice gauge theory,
\index{lattice gauge theory (LGT)}%
 characterises the physical behaviour of the fields locally \emph{at the lattice scale} ignoring correlations over distances much larger than the lattice spacing and long wavelength effects.   The fields correspond to a phase transition between confined and Coulomb-like phases.

\smallskip

The formal gauge symmetry
\index{gauge ! symmetry}%
 of equation (5) is sufficiently close to the physics to produce a Coulomb phase with a massless photon.  To demonstrate this in the mean field approximation,
\index{mean field approximation (MFA)}%
 link and site variables are assumed to have expectation values which are determined by self-consistency conditions, that is one link or site variable is allowed to fluctuate quntum mechanically whilst all fields on neighbouring links and sites are replaced by their mean values.  Working one link or site at a time spontaneously breaks the local gauge symmetry
\index{gauge ! symmetry}%
 without needing to work in a fixed gauge.

Translational and rotational
\index{group ! orthogonal (rotation)}%
 invariance of the vacuum
\index{vacuum}%
 are used to make the gauge-Higgs mean field \index{mean field approximation (MFA)}%
ansatz:
\begin{equation}
\langle U_h (\xymatrix{
{\bullet}_x \ar@{-}[r] & {\bullet}_y) \rangle = V_{U_h}\quad \text{and}\quad  \langle H(x) \rangle = V_H,}
\end{equation}
where $V_{U_h}$ and $V_H$ assume the same real value for each link and site respectively.  From equation (7), the real expectation value $V_{U_h}$ is independent of the link orientation, and because $H^{-1} (x) = H^{*} (x)$ (as for an antiunitary operator) it follows that $ \langle H^{-1}(x) \rangle = V_H$.

To calculate an effective single-link action replace all variables except for $U_h (\xymatrix{ {\bullet}_x \ar@{-}[r] & {\bullet}_y) }$ by their mean value in the action of equation (9).  Then look for the unique self-consistent solutions of $V_{U_h}$ and $V_H$ for every coupling constant pairing $(\alpha, \beta)$, this being the physical solution with the lowest free energy, that is the largest value of $\log Z$ where $Z$ is the partition function
\[ Z = \int DU_h DH\  \text{exp}[\mathcal{S}(U_h, H)]. \]
The 3 phases in the mean field approximation
\index{mean field approximation (MFA)}%
 are:-

1) \emph{Strong coupling or confined phase}, including the region of small $\alpha$ and $\beta$ values and corresponds to the trivial mean field solution with $V_H = 0 = V_{U_h}$.

2) \emph{Higgs phase}, including the region of large $\alpha$ and $\beta$ values and corresponding to a solution with $V_H, V_{U_h} \neq 0$.

3) \emph{Coulomb phase}, including the region of large $\beta$ values and bounded $\alpha$ and corresponding to a solution with $V_H = 0$ but $V_{U_h} \neq 0$.

In the Coulomb phase there are long-range correlations corresponding to a massless photon due to the formal gauge invariance,
\index{gauge ! invariance}%
 that is the gauge properties of the vacuum.
\index{vacuum}%

From equation (8) the vacuum expectation values (VEV)
\index{vacuum expectation value (VEV)}%
 $V_{U_h}$ and $V_H$ under global gauge transformations
\index{gauge ! transformation}%
 with a constant gauge function $\alpha$ behave as:

$\Lambda_G (x) = e^{i\alpha}$: $V_{U_h} \to V_{U_h}$ and $V_H \to e^{-i\alpha} V_H$\\
and under local gauge transformations with a linear gauge function

$\Lambda_L (x) = e^{i\alpha^{\mu} x_{\mu}}$: $V_{U_h} \to e^{i\alpha^{\mu} x_{\mu}} V_{U_h}$ and $V_H \to e^{-i\alpha^{\mu} x_{\mu}} V_H$,\\
where $a_{\mu}$ is the appropriate lattice link vector.
\index{lattice}%

It is then argued~\cite[p.~117]{froggatt} that a phase containing a massless photon must have that (i) the gauge symmetry
\index{gauge ! symmetry}%
 for linear gauge functions be spontaneously broken, that is the vacuum
\index{vacuum}%
 is not invariant under $\Lambda_L (x)$, but that (ii) the global part of the gauge symmetry must not be spontaneously broken, that is the vacuum is invariant under a constant $\Lambda_G(x)$.  Because the Coulomb phase satisfies both $V_{U_h} \neq 0$ and $V_H = 0$, it must contain a massless gauge particle (the photon).
\index{gauge}%

So the dynamics of the gauge non-invariant action of equation (4) for $U(1)$ lattice electrodynamics can generate a massless gauge symmetric theory with no parameter fine-tuning.  

This generalizes to a non-abelian lattice gauge theory
\index{lattice gauge theory (LGT)}%
 with a gauge-symmetry
\index{gauge ! symmetry}%
 violating lattice action of equation (4) which is effected by taking the $U(-)$ link variables to be non-abelian group
\index{group ! non-abelian}%
 valued.  In the non-abelian theory there is no region in $(\alpha, \beta)$ parameter space corresponding to a Coulomb phase, but this appearance of gauge symmetry ``ex nihilo'' also follows for a non-abelian group~\cite[p.~118]{froggatt}.

\medskip

Thus the field theory glass model can thus be used to demonstrate the very general nature of the inverse Higgs mechanism and the spontaneous appearance of gauge symmetry.  A thought experiment to produce a Monte Carlo simulation of such a model begins with a randomly selected set of sites $\{ i \}$ in four-dimensional space-time.  A manifold $M_i$ is constructed at each point $i$ on the random lattice
\index{lattice}%
 and varies from site to site.  The basic dynamical variable is a generalized quantum field $\phi(i) : i \to M_i$.  The parameters of an action $\mathcal{S}$ are randomly chosen numbers. Monte Carlo methods are used to study long wavelength properties.  The claim is that the physical degrees of freedom, with the most long range correlations, are similar to the gauge degrees of freedom in the above translational invariant lattice models. (Translational invariance is also key to the proof of the Nambu-Goldstone theorem).  So the introduction of variables such as $U_h$ and $H$ should be possible, leading to a precise formal gauge invariance, which in turn is realised as massless gauge particles or as low energy confinement.

The non-Poincar\'e invariant field theory glass action is a sum 
\[ \mathcal{S}[\phi] = \sum_r \mathcal{S}_r(\phi(i)),\quad i \in r\]
over quenched random terms from small overlapping space-time regions $r$ of the order of the Planck size.   Each small region of space-time has an approximate local gauge symmetry of the action, together with a gauge group.  Assume that within the totality of space-time there is an overlapping area with a large density of such regions.

The approximate local symmetry groups $G(s)$ for a site $s$ now converts into exact gauge symmetries through the introduction of `by hand' variables $(\phi_h, H)$, where the Higgs field is defined on $s$.  Define $\phi_h$ to have the same structure as the fundamental field $\phi$, by
\[ \phi(i) = \phi_{h}^{H} (i), \]
where $\phi^{\Omega}$ is the mapping of $\phi$ by an element $\Omega$ from the direct product of all the local symmetry groups $G(s)$.  The field $\phi_h (i)$ at site $i$ is transformed only under the $G(s)$ for which the gauge area contains $i$.

The action $\mathcal{S}[\phi]$ expressed in terms of new variables, $\mathcal{S}[\phi_h, H] = \mathcal{S}[\phi = \phi_h^{H}]$ is automatically stable under the formal gauge transformation
\index{gauge ! transformation}%
\[ \phi_h \to \phi_h^{\Omega},\quad H \to \Omega^{-1} H \]
because $\phi$ is also invariant under this artificial map
\[ \phi = \phi_h^{H} \to (\phi_h^{\Omega})^{\Omega^{-1} H} = \phi_h^{H} = \phi. \]

The field theory glass expressed in terms of $(\phi_h, H)$ manifests a formal gauge symmetry in which the symmetry group changes \emph{randomly} from place to place. 

Any given gauge group $K$
\index{group ! gauge}%
 is an approximate gauge symmetry of $\mathcal{S}[\phi]$.  To discuss the situation where the associated Higgs field $H$ in the new action $\mathcal{S}[\phi_h, H]$ has local fluctuations and no long-range correlations, and so a vanishing `mean value' in the vacuum,
\index{vacuum}%
 focus on gauge areas within which the symmetry breaking terms of $\mathcal{S}[\phi]$ are small corresponding to a small $\alpha$ in the previous lattice
\index{lattice}%
 model.  In order to achieve vacuum invariance under global transformations assume that $\langle H \rangle = 0$.

To describe deviations from the globally gauge invariant vacuum
\index{vacuum}%
 state, modify the vacuum state by a continuum Yang-Mills gauge field $A_{\mu}^{a} (x)$ for $K$ with generators $\lambda^a / 2$.  Consider all gauge areas containing $i$ with small-enough $K$-symmetry violating terms in $\mathcal{S}[\phi]$ to ensure that the correlation function between the Higgs field at two points, $H(x)$ and $H(y)$ decays exponentially with the separation.  Assume $A_{\mu}^{a} (x)$ is constant over all the gauge areas containing $i$ and having coordinates $x_i^{\mu}$.  In order to modify $\phi_h$ at each site so as to establish a new configuration corresponding to $A_{\mu}^{a} (x)$, apply to $\phi_h (i)$ one gauge transformation
\index{gauge ! transformation}%
\[ \Lambda_s (x_i) = \text{exp} [ i A_{\mu}^{a} (x_i) \frac{\lambda^{\mu}}{2} ( x_s^{\mu} - x_i^{\mu} ) ] \]
per gauge area containing $x_i$.  The $x_s^{\mu}$ are coordinates of the gauge area centre $s$.  The gauge transformation $\Lambda_s (x_i)$ has the same dependence on gauge area center $x_s^{\mu}$ for every field variable $\phi_h(i)$.  

The above construction of a continuum gauge field fails for a group with outer automorphisms
\index{automorphism ! outer}%
 but can be salvaged if the field theory glass representations are not symmetric under the automorphism.  Whilst the SM group  $\S(\U(2) \times \U(3))$
\index{group ! S(U(2) \times U(3))@$\S(\U(2) \times \U(3))$}%
 is symmetric under complex conjugation (see next chapter), the chiral quark and lepton representations are not.

For each continuum field configuration $A_{\mu}^{a} (x)$ assign a new configuration of the field theory glass.  The natural correspondence between gauge transformations
\index{gauge ! transformation}%
 on the $A_{\mu}^{a} (x)$ and the field theory glass variables $\phi_h(i)$ means that the effective action for $A_{\mu}^{a} (x)$ is invariant under $K$ gauge group
\index{group ! gauge}%
 transformations.  Thus if $K$ has no outer automorphisms
\index{automorphism ! outer}%
 a Yang-Mills field theory can be established for it, and if it does have outer automorphisms
\index{automorphism ! outer}%
 then arbitrary choices would need to be made all over space-time.

In conclusion the formal gauge invariance arises from the definition of what Froggatt and Nielsen
\index{Froggatt, C. D.}%
\index{Nielsen, H. B.}%
 call the `human variables' $(\phi_h, H)$.

\medskip

Importantly for potential applications of our connections between random graphs
\index{graph ! random ($\mathfrak{R}$)}%
 and lattices,
\index{lattice}%
 the lattice used in the above derivations do not have to be regular but can just as well be a discretized model, which in Random Dynamics
\index{Random Dynamics}%
 can be taken to be fixed sites distributed randomly in four-dimensional space-time.  Equally this mechanism should work for very general field theories including a ``field theory glass '' where coupling constants, the type and number of degrees of freedom and other parameters are chosen in a \emph{random quenched way}, that is, fixed and not varied in the quantum mechanical functional integral - by analogy with the atomic binding in a real glass.

From a philosophical point of view, if everything were totally random, then it would be hard to discern much physics; it is because there are some constants or patterns despite the randomness that we can pinpoint physical laws.  Some of these could be the quenched random terms that have been identified in the Random Dynamics program.

\bigskip
\bigskip
\bigskip
\bigskip

\emph{The Multiple Point Principle (MPP)}
\index{Multiple Point Principle (MPP)}%

We saw how the continuum gauge field $A^{a}_{\mu} (x)$ can be implemented on the gauge glass by a suitable modification of the gauge glass degrees of freedom $\phi_h (i)$.   If the gauge field $A^{a}_{\mu} (x)$ for the gauge group $K$ can only be implemented very sparsely in space-time rather than throughout space-time, then the effective continuum gauge field Lagrangian density acquires a term of the form
\[ \mathcal{L}_{\text{eff}} = - \frac{1}{4 g^2} F^a_{\mu\nu} (x) F^{\mu\nu a} (x) \] 
with a very small coefficient $- \frac{1}{4 g^2}$ because $A^{a}_{\mu} (x)$ would represent relatively few gauge glass degrees of freedom.  This implies a strong gauge coupling constant $g^2$, so the gauge degrees of freedom are confined close to the Planck scale and become irrelevant to the low energy levels accessible in the laboratory.  For a gauge group $K$ to survive to low energies it must be a good approximate symmetry of the fundamental field theory glass action $S[\phi]$.

\medskip

One of the assumptions of the Random Dynamics
\index{Random Dynamics}%
 project is that there is an \emph{anti-unified}
\index{anti-unification}%
 gauge group
\index{group ! gauge}%
 $\S(\U(2) \times \U(3))^{N_{gen}}$
\index{group ! S(U(2) \times U(3))@$\S(\U(2) \times \U(3))$}%
 (see below) on which the plaquette action of the lattice gauge theory
\index{lattice gauge theory (LGT)}%
 is defined as a direct product of $N_{gen} = 3$ Standard Model
\index{Standard Model (SM)}%
 group factors of the form $\SU(3)^{N_{gen}} \times \SU(2)^{N_{gen}} \times \U(1)^{N_{gen}}$, 
 \index{group ! SU(3) \times SU(2) \times U(1)@$\SU(3) \times \SU(2) \times \U(1)$}%
 that at the Planck scale ($1.22 \times 10^{19}$ GeV) breaks down to the \emph{diagonal subgroup} $\S(\U(2) \times \U(3))_{diag} := \{( g, g, g) : g \in \S(\U(2) \times \U(3)) \} \cong \S(\U(2) \times \U(3))$
\index{group ! S(U(2) \times U(3))@$\S(\U(2) \times \U(3))$}%
 due to a so-called \emph{confusion mechanism}~\cite{benett4}~\cite{bennettnielsen}
\index{confusion mechanism}%
 arising from group outer
\index{automorphism ! outer}%
 automorphisms - which of the identical but different groups being acted on by the outer automorphism
\index{automorphism ! outer}%
 to choose.  The diagonal subgroup consists of ordered sets of SM group elements $(g, g, \ldots, g)$ for which all the components are equal to each other.  In fact the speculation is that there is a successive breaking of groups with many direct products down to a collection of groups with particularly few automorphisms such as $\S(\U(2) \times \U(3))$
\index{group ! S(U(2) \times U(3))@$\S(\U(2) \times \U(3))$}%
 itself, with $\S(\U(2) \times \U(3))^3$ being the penultimate group in this series.

Given the diagonal subgroup, the naive lattice continuum formula on an assumed fundamental lattice with lattice constant $a$ is 
\[ U(\xymatrix{{\bullet}_{x_{\mu}} \ar@{-}[r] & {\bullet}_{y_{\mu}})} \simeq exp(i g a A^{a}_{\mu} \frac{1}{2} \lambda^a),\quad y_{\mu} = x_{\mu} + a \delta_{\mu},\] 
implies
\[(g A^{a}_{\mu})_1 = (g A^{a}_{\mu})_2 = \ldots = (g A^{a}_{\mu})_{N_{gen}}.\]

For such a diagonal subgroup, the original fundamental Lagrangian density
\[ \mathcal{L} = - \sum_{i=1}^{N_{gen}} \frac{1}{4 g^{2}_i}  (g F^a_{\mu\nu})^{2}_i  \]
becomes
\[ \mathcal{L} = - \frac{1}{4 g^{2}_{\text{diag}}}  (g F^a_{\mu\nu})^{2}_{\text{diag}}, \]
with
\[ \frac{1}{4 g^{2}_{\text{diag}}}  =  \sum_{i=1}^{N_{gen}} \frac{1}{4 g^{2}_{i}}.   \]
\medskip

If the couplings for the `fundamental' fields $(g A^{a}_{\mu})_i$ are all assumed at the fundamental scale to be equal to a certain critical coupling at that scale then the observed coupling will obey
\[ \frac{1}{g^{2}_{\text{diag}}} = \frac{1}{g^{2}_{\text{crit}}} N_{gen}. \]

\medskip
If many vacua are degenerate then the corresponding phases meet at a certain point in the phase diagram of the theory.  The phase diagram of any gauge theory is represented by a space which has axes given by bare coupling constants, and perhaps also by bare masses. 

The criticality of the gauge coupling constants at the Planck scale means that at this scale the couplings lie at the boundary between the Coulomb (photon) and the strong coupling (confining) phases.  A calculation of critical couplings at a phase boundary, applying to both phases, can be done using the mean field approximation (MFA).
\index{mean field approximation (MFA)}%

\bigskip

The MPP mechanism within the Random Dynamics
\index{Random Dynamics}%
 program postulates the existence of a point in the phase diagram for a Yang-Mills lattice gauge theory
\index{lattice gauge theory (LGT)}%
 of the Standard Model
\index{Standard Model (SM)}%
 with gauge group
\index{group ! gauge}%
 $\S(\U(2) \times \U(3))^3$
\index{group ! S(U(2) \times U(3))@$\S(\U(2) \times \U(3))$}%
 where a maximum number of vacuum
\index{vacuum}%
 phases (all with the same energy density) come together so that the vacuum is maximally degenerate. The Principle of Multiple Point Criticality
\index{Principle of Multiple Point Criticality}%
 then requires that the running gauge
\index{gauge}%
 coupling constants correspond to the multiple point critical values in the phase diagram of a lattice theory and so the MPP explains the fine tuning of the free parameter values of the Standard Model.  The multiple point is where all or at least a maximum number of the phases of the lattice gauge theory
\index{lattice gauge theory (LGT)}%
 meet.  `Phase' here means different behaviour, analogous to the 3 phases of water; in the relevant phase diagram water and vapour are indistinguishable at the critical point and ice, water and vapour exist simultaneoulsy at the triple point.    The Multiple Point Principle was born in the lattice investigations of gauge theories, and particularly Monte Carlo simulations of $\U(1)$, $SU(2)$ and $SU(3)$
\index{group ! SU(2)@$\SU(2)$}%
\index{group ! SU(3)@$\SU(3)$}%
\index{group ! U(1)@$\U(1)$}%
 gauge theories on the lattice which alluded to the existence of the triple critical point.

There is a phase for each normal (invariant) subgroup
\index{group ! normal (invariant) subgroup}%
 of $\S(\U(2) \times \U(3))^3$.
\index{group ! S(U(2) \times U(3))@$\S(\U(2) \times \U(3))$}%
  The system is very stable because there is an entire range of parameter values for say the mean energy and volume per molecule.  The observed coupling values are the critical multiple point values that the couplings take when the vacuum
\index{vacuum}%
 is maximally degenerate.

The multiple point values of the plaquette action parameters for the diagonal subgroup of 
\index{group ! S(U(2) \times U(3))@$\S(\U(2) \times \U(3))$}%
$\S(\U(2) \times \U(3))^3$ are in the continuum limit equal to the experimental gauge
\index{gauge}%
 coupling values that have been extrapolated to the Planck scale, thereby justifying use of the gauge group
\index{group ! gauge}%
 $\S(\U(2) \times \U(3))^3$.
\index{group ! S(U(2) \times U(3))@$\S(\U(2) \times \U(3))$}%

Grand Unified Theory models based on simple gauge
\index{group ! gauge}%
 groups $\SU(5)$
\index{group ! SU(5)@$\SU(5)$}%
 and $\SO(10)$
\index{group ! SO(u10)@$\SO(10)$}%
 place left-handed fermions and their antiparticles in the same representation and seem to lead to families with a pair of almost mass degenerate particles.  But inter-generation masses are not approximately degenerate~\cite[p.~171]{froggatt}.

Random Dynamics
\index{Random Dynamics}%
bypasses considerations of such gauge groups, and further assumes that the chiral horizontal symmetry group commutes with the SM gauge group.
\index{group ! gauge}%
  Random Dynamics, through the confusion mechanism
\index{confusion mechanism}%
 breakdown, does seem to favour the SM gauge group $\S(\U(2) \times \U(3))$,
\index{group ! S(U(2) \times U(3))@$\S(\U(2) \times \U(3))$}%
 and furthermore relates the number 3 of generations to the phenomenological values of the gauge coupling constants for the 3 invariant subgroups of the SM group~\cite{bennettniel}, but this result essentially requires there to be no new phenomena between accessible energies and that of the Planck scale.

One of the most important conclusions of Das and Laperashvili's review~\cite{das}
\index{Das, C. R.}%
\index{Laperashvili, L. V.}%
of the Multiple Point Principle (MPP)
\index{Multiple Point Principle (MPP)}%
 is the validity of an approximate universality of the critical couplings. They showed that values of the phase transition couplings can be crudely approximated without using any specific lattice, so that the details of both the lattice and the regularization are unimportant.  Critical couplings depend only on groups with some regularization.  This approximate universality is essential in comparisons of the lattice phase transition couplings with the experimentally observed couplings.

\medskip

Within the Grand Unified Theories program it is assumed that different irreducible representations (IR)
\index{irreducible representations}%
 of the $\S(\U(2) \times \U(3))$
\index{group ! S(U(2) \times U(3))@$\S(\U(2) \times \U(3))$}%
 will combine into a single IR of a larger group, however this approach does not seem to produce a satisfactory explanation of the three family generation structure within the SM.   Family (or horizontal) symmetries are to be differentiated from GUT symmetries which unify different members within a family.  It was argued in~\cite{froggattlowe} that the fermion mass hierarchy suggests extensions of the SM which have the same IR as arise in the SM and which embed $\S(\U(2) \times \U(3))$ as a diagonal subgroup.  The fact that no pair of correponding particles in different generations is degenerate makes it less likely that horizontal flavour groups can account for the generation gaps: if such pairs were positioned into multiplets of groups such as $\SU(2)_{H}$
\index{group ! SU(2)@$\SU(2)$}%
 in an $S(\U(2) \times \U(3)) \times \SU(2)_{H}$
\index{group ! S(U(2) \times U(3))@$\S(\U(2) \times \U(3))$}%
 model it would suggest that these particles are degenerate.  Also in order for an over-group $G$ of $\S(\U(2) \times \U(3))$
\index{group ! S(U(2) \times U(3))@$\S(\U(2) \times \U(3))$}%
 to account for the gaps, corresponding particles from different generations should belong to inequivalent IRs of $G$, and furthermore single IRs of the $\S(\U(2) \times \U(3))$
\index{group ! S(U(2) \times U(3))@$\S(\U(2) \times \U(3))$}%
 should form IRs of $G$.  A more interesting possibility is the diagonal embedding suggested above.

Whilst the SM does not contain any quantum number which distinguishes the lepton and quark generations it is suspected that such a number must exist, and the 3-coloured random graph
\index{graph ! triality ($\mathfrak{R^{t}}$)}%
 provides one approach to finding it.  Having $N_{gen}$ many $S(\U(2) \times \U(3))$
\index{group ! S(U(2) \times U(3))@$\S(\U(2) \times \U(3))$}%
 factors allows for different gauge
\index{gauge}%
 quantum numbers for each of the $N_{gen}$ families~\cite{bennettnielsen}, an idea that mirrors the possibility that we have suggested that these are colours of the triality graph~\cite{tarzi2}.
\index{graph ! triality ($\mathfrak{R^{t}}$)}%
  We return to this point in the next chapter.

\smallskip

\emph{Tight Packing}

For all of the successes of the Standard Model
\index{Standard Model (SM)}%
 it predicts neither the number of generations nor the pattern of generation replication.  It also leaves unresolved the large number of free parameters.  A Random Dynamics-inspired
\index{Random Dynamics}%
 model relating the smallness of the gauge coupling constants to the number of generations predicts 3 generations~\cite{bennettniel}.

In~\cite{tarzi2} it was suggested that a tight packing interpretation of the leptons of the Standard Model
\index{Standard Model (SM)}%
 in some space could be part of the solution to the family structure within the SM.  A. Kleppe
\index{Kleppe, A.}%
 has used a tightest packed three-dimensional lattice
\index{lattice}%
 which basically represents the abelian sector of the Standard Model
\index{Standard Model (SM)}%
 group, i.e. $\U(1)^3$
\index{group ! U(1)@$\U(1)$}%
 in an $N_{gen} = 3$-dimensional space corresponding to three generations, within the anti-grand unification
\index{anti-unification}%
 scheme.  The tight packing arises because each lattice site has a maximal number of neighbouring sites.

Dense packing means that each lattice site has the maximum number of nearest neighbours at the same critical distance. 

Each lattice direction where the nearest neighbour distance is smaller than that corresponding to the critical coupling value, has a confining subgroup.  The critical coupling is usually a function of the remaining $\U(1)$
\index{group ! U(1)@$\U(1)$}%
 degrees of freedom.  But the critical distance in one direction is approximately independent of the nearest neighbour distance in other directions, so approximate MPP criticality is achieved when the nearest neighbour
distance is critical in all directions.  This means a maximal number of $1$-dimensional subgroups, or having the tightest possible packing of lattice
\index{lattice}%
 points.  This is the origin of tight packing in Kleppe's
\index{Kleppe, A.}%
 argument.  The lattice packing density is the ratio of the volume of a sphere to the volume of the fundamental region.

In lattice packing a regular sphere is a repetition of one lattice
\index{lattice}%
 crystal but in sphere-packing the position of every sphere is a variable.

The origin and nature of the tight packing differs in the random graph
\index{graph ! random ($\mathfrak{R}$)}%
 context and the lattice gauge theory 
\index{lattice gauge theory (LGT)}%
context, but it is interesting that the same general idea has been called upon.

\chapter{Outer Automorphisms}
\index{automorphism ! outer}%
\label{oaut}

The groups $\U(1)$,
\index{group ! U(1)@$\U(1)$}%
 $\SU(n)$ $(n > 2)$
\index{group ! SU(n)@$\SU(n)$}%
 and the SM group $\S(\U(2) \times \U(3))$
\index{group ! S(U(2) \times U(3))@$\S(\U(2) \times \U(3))$}%
 have an outer automorphism
\index{automorphism ! outer}%
 that corresponds to charge conjugation symmetry.  For $\SU(n)$
\index{group ! SU(n)@$\SU(n)$}%
 groups ($n > 2$) it is complex conjugation of the matrix elements.   In the SM this symmetry is broken because only left-handed fermions and right-handed anti-fermions couple to the $W^{\pm}$ intermediate vector boson.  Note that $\SU(2),
\index{group ! SU(2)@$\SU(2)$}%
\index{group ! SO(3)@$\SO(3)$}%
 \SO(3)$, odd-$n$ spin and symplectic groups
\index{group ! spin}%
\index{group ! symplectic}%
 have no outer
\index{automorphism ! outer}%
 automorphisms.

Chirality breaks the charge conjugation symmetry of the gauge fields and ensures that gauge groups
\index{group ! gauge}%
 such as $\U(1)$
\index{group ! U(1)@$\U(1)$}%
 and $\SU(3)$
\index{group ! SU(3)@$\SU(3)$}%
 survive the breakdown from larger to smaller groups.   The confusion
\index{confusion mechanism}%
 breakdown through a charge conjugation 
\index{automorphism}%
automorphism is not possible because left- and right-handed fermions appear in different representations.  

However there may be non-representation theoretic reasons for a phenomenon.  The order-2 outer
\index{automorphism ! outer}%
 automorphism of $\Aut(\mathfrak{R})$
\index{group ! Aut(\mathfrak{R})@$\Aut(\mathfrak{R})$}%
 might be taken to correspond to charge conjugation~\cite{tarzi2}.

The Random Dynamics
\index{Random Dynamics}%
 program goes further assuming that a gauge field can be permuted with an automorphic image of itself, and that this can happen locally and not globally.  The diagonal subgroup of the direct product of isomorphic groups is the maximal invariant subgroup of the permutation
\index{automorphism}%
 automorphisms;  it is symmetric under gauge transformations
\index{gauge ! transformation}%
 generated by constant gauge functions that correspond to the global part of a local gauge transformation.  

The diagonal subgroup of $\S(\U(2) \times \U(3))^{N_{gen}}$
\index{group ! S(U(2) \times U(3))@$\S(\U(2) \times \U(3))$}%
 is stabilized by the automorphic permutation of the $N_{gen}$
\index{group ! S(U(2) \times U(3))@$\S(\U(2) \times \U(3))$}%
 factors in $\S(\U(2) \times \U(3))^{N_{gen}}$.
\index{group ! S(U(2) \times U(3))@$\S(\U(2) \times \U(3))$}%
  But then the matter field content of each factor must have the same structure.  This and the known fact that what is normally taken to be a fermion generation gives the simplest arrangement of particles that avoids gauge anomalies suggests that the $N_{gen}$ factors of $\S(\U(2) \times \U(3))^{N_{gen}}$
\index{group ! S(U(2) \times U(3))@$\S(\U(2) \times \U(3))$}%
 are just repetitions of the $\S(\U(2) \times U(3))$,
\index{group ! S(U(2) \times U(3))@$\S(\U(2) \times \U(3))$}%
 one for each generation.

The assumption is that the final breakdown of gauge symmetry by confusion
\index{confusion mechanism}%
 is caused by the automorphism
\index{automorphism}%
 that permutes the $N_{gen}$ isomorphic product factors $\S(\U(2) \times \U(3))$ in $\S(\U(2) \times \U(3))^{N_{gen}}$.
\index{group ! S(U(2) \times U(3))@$\S(\U(2) \times \U(3))$}%

Random Dynamics
\index{Random Dynamics}%
 suggests the existence of an intermediate gauge group
\index{group ! gauge}%
$\S(\U(2) \times \U(3))_1 \times  \S(\U(2) \times \U(3))_2 \times  \S(\U(2) \times \U(3))_3$
at the confusion
\index{group ! S(U(2) \times U(3))@$\S(\U(2) \times \U(3))$}%
\index{confusion mechanism}%
 scale; with one direct product factor $\S(\U(2) \times \U(3))_i$ for each quark-lepton generation.  With such a direct product unified gauge group,
\index{group ! gauge}%
 it is possible that the separate gauge quantum numbers of $\S(\U(2) \times \U(3))_1$, $\S(\U(2) \times \U(3))_2$ and $\S(\U(2) \times \U(3))_3$
\index{group ! S(U(2) \times U(3))@$\S(\U(2) \times \U(3))$}%
 could play the role of horizontal quantum numbers and be responsible for the different mass scales of the three generations.  But what is the origin of these quantum numbers?  Even within Random Dynamics
\index{Random Dynamics}%
 there is the problem of understanding the origin of the $\S(\U(2) \times \U(3))$
\index{group ! S(U(2) \times U(3))@$\S(\U(2) \times \U(3))$}%
 and of the number of quark and lepton generations.  We outlined a different type of solution based on the 3-coloured random graph in~\cite{tarzi2}.
\index{graph ! triality ($\mathfrak{R^{t}}$)}%

Random Dynamics
\index{Random Dynamics}%
relates the number $N_{gen} = 3$ of generations to the phenomenological values of the gauge coupling constants for the three invariant subgroups of $\S(\U(2) \times \U(3))$,
\index{group ! S(U(2) \times U(3))@$\S(\U(2) \times \U(3))$}%
 a result that seems to require there to be no new physics between presently accessible energies and the Planck scale, which is surely unlikely.

There is an additivity of the inverse squared couplings of the group factors in the direct product $\S(\U(2) \times \U(3))^3$
\index{group ! S(U(2) \times U(3))@$\S(\U(2) \times \U(3))$}%
 in going to the diagonal subgroup of the form
\[ \frac{1}{g^{2}_{i, diag}}  =  \sum_{j=1}^{N_{gen}} \frac{1}{g^{2}_{i, j}} = \frac{3}{g^{2}_{MP}}\quad (i \in \{\SU(2), \SU(3) \}), \]
that is at the multiple point
\index{Multiple Point Principle (MPP)}%
 each term in this series become equal to the same value.  For the abelian $\U(1)$
\index{group ! abelian}%
\index{group ! U(1)@$\U(1)$}%
  the deviation of the coupling in going from the multiple point of $\S(\U(2) \times \U(3))^3$
\index{group ! S(U(2) \times U(3))@$\S(\U(2) \times \U(3))$}%
 to the diagonal subgroup is 6 instead of 3. 

When considering the action corresponding to $\S(\U(2) \times \U(3))^{3}$
\index{group ! S(U(2) \times U(3))@$\S(\U(2) \times \U(3))$}%
 the abelian
\index{group ! abelian}%
 part is treated separately from the non-abelian part.  For the non-abelian subgroups the action is additive $\mathcal{S} = \sum_j \mathcal{S}_j$ meaning the confining phases correspond to factorizable normal (invariant)
\index{group ! normal (invariant) subgroup}%
 subgroups so the phase diagram for $\S(\U(2) \times \U(3))^{3}$
\index{group ! S(U(2) \times U(3))@$\S(\U(2) \times \U(3))$}%
 is determined from those of the individual factors.

The $3$-dimensionality of the abelian group $\U(1)^3$
\index{group ! abelian}%
\index{group ! U(1)@$\U(1)$}%
 gives the underlying lattice
\index{lattice}%
 and the number of generations.  This lattice brings together a maximal number of phases and is found using the Principle of Multiple Point Criticality.
\index{Multiple Point Principle (MPP)}%
  The lattice sites can correspond to quantized quantities such as charges.  To introduce dynamics into this picture absorb the $\U(1)$
\index{group ! U(1)@$\U(1)$}%
 coupling into the metric $g_{\mu \nu}$ on the space of charges.  Taking $\U(1)$ to be $e^{i\theta} \sim e^{i g A^{\mu}}$, the metric is used in forming the inner product of two $\theta$-variables.  The $\U(1)$
\index{group ! U(1)@$\U(1)$}%
 viewpoint is that the lattice
\index{lattice}%
 sites $0, 2\pi, 4\pi, \ldots 2n\pi\pmod 2$ are identified with the origin.  But by including the coupling into the definition of distance, the distance between two charges will be a function of the values of the charges.

For confinement along $\U(1)_j$
\index{group ! U(1)@$\U(1)$}%
 the critical distance is longer than the lattice link distance.
  For every pair $(G_i, N_i)$ of subgroups of $\S(\U(2) \times \U(3))$,
\index{group ! S(U(2) \times U(3))@$\S(\U(2) \times \U(3))$}%
 where $N_i$ is a normal
\index{group ! normal (invariant) subgroup}%
 subgroup of $G_i$, there is a phase, and all the phases contribute to the action.  For example, $(U_1, \mathbb{Z}_2)$ begets $\mathcal{S}_{\square} = \beta cos(\theta)$ as a lattice action~\cite{kleppe}.

\bigskip

In every small space-time region there is an ambiguity in choosing how to represent the continuum gauge field.  The difficulty in implementing a continuum Yang-Mills field $A^{a}_{\mu} (x)$ on a gauge glass
\index{gauge ! glass}%
 with gauge group
\index{group ! gauge}%
 $K$ having outer
\index{automorphism ! outer}%
 automorphisms, is in deciding which degrees of freedom of the gauge glass
\index{gauge ! glass}%
 to identify with which degrees of freedom of $A^{a}_{\mu} (x)$~\cite[pp.~152-173]{froggatt}.

By contrast the the ambiguity in identifying the continuum field due to inner
\index{automorphism ! inner}%
 automorphisms is just the usual gauge ambiguity in defining the gauge field.  It adds a background field but does not give rise to any ambiguity in the physics.  For an inner
\index{automorphism ! inner}%
 automorphism, 
\[ f(g) = b g b^{-1},\quad b, g \in K, \]
the transformed field $A^{fa}_{\mu} (x)$ is related to $A^{a}_{\mu} (x)$ by a gauge transformation,
\index{gauge ! transformation}%
 with a constant gauge function:
\[ \frac{\lambda^a}{2} A^{fa}_{\mu} (x) = U(x) \frac{\lambda^a}{2} A^{a}_{\mu} (x) U^{-1} (x) - i \partial_{\mu} U(x) U^{-1} (x) \]
where $U(x) = b$ and $\frac{\lambda^a}{2}$ are group generators.  

Inconsistencies corresponding to the inner
\index{automorphism ! inner}%
 automorphism $f(g) = b g b^{-1}$ can be gauge transformed away but not those corresponding to outer automorphisms. 
\index{automorphism ! outer}%

Because a continuum Yang-Mills theory for a gauge group
\index{group ! gauge}%
 with outer
\index{automorphism ! outer}%
 automorphisms is unlikely to be implemented on a gauge glass,
\index{gauge ! glass}%
 Random Dynamics
\index{Random Dynamics}%
 predicts that the continuum gauge group
\index{group ! gauge}%
 must have no outer
\index{automorphism ! outer}%
 automorphisms that can be extended to true discrete symmetries. 

\bigskip

\emph{Generalised outer automorphisms}
\index{automorphism ! outer}%
 of a non-semisimple group
\index{group ! non-semisimple}%
 are isomorphisms between two of its factor groups obtained by dividing out low order normal
\index{group ! normal (invariant) subgroup}%
 subgroups.  They act as ordinary outer
\index{automorphism ! outer}%
 automorphisms of the Lie algebra
\index{Lie algebra}%
 and thus of the continuum gauge fields $A^{a}_{\mu} (x)$.

According to the Random Dynamics
\index{Random Dynamics}%
 program a generalised outer
\index{automorphism ! outer}%
 automorphism causes an inconsistency in the identification convention for the gauge glass
\index{gauge ! glass}%
 degrees of freedom, because of an ambiguity involving the identification of a factor group of the gauge group
\index{group ! gauge}%
 $K$ of the continuum gauge field.  Such an identification means that the group elements in the coset of the factor group must correspond to the same continuum field $A^{a}_{\mu} (x)$.  There should then be an approximate symmetry under permutations of the the group elements inside a coset.

What Froggatt and Nielsen
\index{Froggatt, C. D.}%
\index{Nielsen, H. B.}%
 have termed the `confusion mechanism'
\index{confusion mechanism}%
 is supposed to collapse the gauge group
\index{group ! gauge}%
 $K$ with its non-trivial outer
\index{automorphism ! outer}%
 automorphisms near the Planck scale.

Using a generalized lattice gauge theory
\index{lattice gauge theory (LGT)}%
 helps to identify collapse mechanisms that break $K$ somewhere between the Planck scale and the low energy scales accessible to us today.  The following modifications to the lattice gauge theory,
\index{lattice gauge theory (LGT)}%
 listed in~\cite{froggatt} are chosen to simulate the action for a gauge glass
\index{gauge ! glass}%
 but the Feynman path integral
\index{path integral}%
 is stabilised:

1)  Taking the lattice
\index{lattice}%
 sites to assume random positions separated by distances of the order of the Planck length gives an amorphous `glass-like' structure instead of a regular hypercubic lattice. 

2)  The gauge group
\index{group ! gauge}%
 can vary from site to site.

3)  The plaquette variables $U_{\square}$ are formed from link variables for the gauge group
\index{group ! gauge}%
 $K$, denoted $U(\xymatrix{{\bullet} \ar@{-}[r] & {\bullet})}$, as is the plaquette action $\mathcal{S}_{\square}$.  The plaquette variables and action contributions are invariant under the `confused gauge transformation'~\cite[p.~156]{froggatt}:
\index{gauge ! transformation}%

$U(\xymatrix{
{\bullet}_x \ar@{-}[r] & {\bullet}_y) \to f_{x,y}(\Lambda(x))^{-1} }$$U(\xymatrix{{\bullet}_x \ar@{-}[r] & {\bullet}_y) f_{y,x}(\Lambda(y)) }$\quad (\dagger)\\
rather than simply under the usual gauge transformation

$U(\xymatrix{
{\bullet}_x \ar@{-}[r] & {\bullet}_y) \to \Lambda(x)^{-1} }$$U(\xymatrix{{\bullet}_x \ar@{-}[r] & {\bullet}_y) \Lambda(y) }$\\
where the $f_{x,y}$ are quenched randomly chosen outer
\index{automorphism ! outer}%
 automorphisms of the gauge group
\index{group ! gauge}%
 $K$.

4)  Assume that $\mathcal{S}_{\square}$ is a random function of $U_{\square}$.  Gauge invariance requires $\mathcal{S}_{\square}$ to take the same value for all conjugate group elements $U_{\square}$ (those related by an inner automorphism).
\index{automorphism ! inner}%

So $\mathcal{S}_{\square}$ can be expressed as a sum of the characters for the various representations $r$ of $K$,
\[ \mathcal{S}_{\square} = \sum_r \beta_r  \chi_r(U_{\square}), \]
where the coefficients $\beta_r$ are quenched random variables obeying $\beta_r = \beta^{*}_{\overline{r}}$ and $\overline{r}$ denotes the complex conjugate representation to $r$.  The character for representation $r$ is $\chi_r (U_{\square}) = Tr\{\rho_r (U_{\square})\}$ where $\rho_r(U_{\square})$ is the representation matrix for $U_{\square}$ in representation $r$.  The $\beta_r$ coefficients are chosen from a Gaussian distribution, independently for each plaquette and for each pair of conjugate representations.  For $S_{\square}$ to be convergent and reasonably smooth, choose the distribution to have zero mean and a width which decreases with the dimension of the representation.

5)  The link variables are group-valued and the group structure varies randomly from link to link, as in Fig. 7.3 of~\cite[p.~158]{froggatt}.  The plaquette variables $U_{\square}$ are then constructed from common factor groups of the groups on the surrounding links.  Ambiguities arising from outer
\index{automorphism ! outer}%
 automorphisms are inherent in identifying factor groups of different but isomorphic link groups leading us to consider the confused gauge transformations
\index{gauge ! transformation}%
 of point (3).

\medskip

How does the amorphous nature (that is randomness) of these modifications of the usual lattice gauge theory
\index{lattice gauge theory (LGT)}%
 break, at the Planck scale, the gauge groups
\index{group ! gauge}%
 present \emph{a priori} in this gauge glass?
\index{gauge ! glass}%

When the vacuum
\index{vacuum}%
 state is non-invariant under the global gauge symmetry, spontaneous symmetry breakdown occurs, as happens for example when gauge bosons acquire a mass according to the Englert-Brout-Higgs-Guralnik-Hagen-Kibble mechanism.
\index{Englert-Brout-Higgs-Guralnik-Hagen-Kibble (EBHGHK) mechanism}
 The vacuum
\index{vacuum}%
 state of the theory in the classical approximation can be studied by considering individually the energetically preferred value of each plaquette variable $U_{\square}$.  

The link and plaquette variables in the vacuum
\index{vacuum}%
 state must belong to the center
\index{group ! center}%
 of the gauge group
\index{group ! gauge}%
 $K$ otherwise the vacuum
\index{vacuum}%
 will not be invariant under global gauge transformations
\index{gauge ! transformation}%
 and gauge symmetry will collapse.  The following are required to prevent such a collapse:

(A) \emph{Absence of outer automorphisms.}  
\index{automorphism ! outer}%
For a gauge group
\index{group ! gauge}%
 $K$ with non-trivial outer
\index{automorphism ! outer}%
 automorphisms the gauge glass
\index{gauge ! glass}%
 action is invariant under the above confused transformation $(\dagger)$.

Assuming that the vacuum
\index{vacuum}%
 values of the corresponding confused link variables $U(\xymatrix{
{\bullet}_x \ar@{-}[r] & {\bullet}_y)  }$ lie in the center of the group,
\index{group ! center}%
 the vacuum
\index{vacuum}%
 values will only be invariant under global gauge transformations
\index{gauge ! transformation}%
 $\Lambda(x) = \Lambda$ if they obey
\[ f_{x,y}(\Lambda) = \Lambda \]
for all outer
\index{automorphism ! outer}%
 automorphisms $f_{x,y}$ randomly chosen throughout the lattice.
\index{lattice}%
  So the vacuum
\index{vacuum}%
 state is only symmetric under the subgroup which is left invariant by all the outer
\index{automorphism ! outer}%
 automorphisms of $K$, and so $K$ spontaneously breaks down to a subgroup with no or few outer automorphisms.
\index{automorphism ! outer}%

Inner automorphisms
\index{automorphism ! inner}%
 may be transformed into background gauge fields which may in turn cause the gauge particles to gain masses through the Englert-Brout-Higgs-Guralnik-Hagen-Kibble mechanism.
\index{Englert-Brout-Higgs-Guralnik-Hagen-Kibble (EBHGHK) mechanism}
The confusion mechanism
\index{confusion mechanism}%
 boils down to inconsistencies in identification for the gauge glass degrees of freedom, and it is possible to circumvent the mechanism and allow a gauge group
\index{group ! gauge}%
 with a few outer
\index{automorphism ! outer}%
 automorphisms to survive, by including matter degrees of freedom.

(B)  \emph{A non-trivial center.}
\index{group ! center}%
For a vacuum
\index{vacuum}%
 state to be invariant under the global gauge transformations
\index{gauge ! transformation}%
 of a group $K$ with a trivial center,
\index{group ! center}%
 all the plaquette variables $U_{\square}$ must equal the identity element in $K$.  The coefficients of the gauge plaquette action $\mathcal{S}_{\square} = \sum_r \beta_r Tr \{ \rho_r (U_{\square}) \} $ may have either sign, corresponding either to an energy maximum or minimum.  So in about half the cases of the plaquette energy being minimized, the group element will not be central meaning that the vacuum
\index{vacuum}%
 is not invariant under the global $K$-action, which is thus broken.  Therefore in a gauge glass,
\index{gauge ! glass}%
 groups with a trivial center are likely to collapse and groups with a non-trivial center survive.

(C)  \emph{A connected center.}
\index{group ! center}%
 Applying a similar argument to (B) to a set of neighbouring plaquettes instead of just one, suggests that gauge groups
\index{group ! gauge}%
 with a topologically disconnected center tend to collapse.  The plaquettes forming the surface of a generalised 3-cube of the amorphous lattice
\index{lattice}%
 obey the Bianchi identity,
\begin{equation}
\prod_{\square \in\ \mbox{\mancube}} U_{\square} = I.
\label{bianchi}
\end{equation}
which arises from the fact that each link is contained in two plaquettes of the cell and the plaquette variables $U_{\square}$ are thus not independent.  If the $U_{\square}$ lie in the center of the group $K$ then the product of the plaquette variables forming the cell surface must equal the unit element.

The identity (\ref{bianchi}) is not automatically satisfied when each plaquette variable is chosen to minimise its energy.  The energetically preferred individual plaquette variables $U_{\square \text{pref}}$ must belong to the center to stop collapse of $K$, but due to the plaquette action being random we expect the product
\[ F = \prod_{\square \in\ \mbox{\mancube}} U_{\square \text{pref}} \]
to be a random central element.  If $F \neq I$ then the true vacuum
\index{vacuum}%
 values $U_{\square \text{vac}}$ of the plaquette variables cannot equal $U_{\square \text{pref}}$.

If the group center
\index{group ! center}%
 were not topologically connected then for some cells $F$ would belong to a different connected component to that of the unit element.  It follows that some of the deviations $U_{\square \text{pref}}\ U_{\square \text{vac}}^{-1}$ would be non-central, so as the $U_{\square \text{pref}}$ are assumed to be central, some of the $U_{\square \text{vac}}$ are non-central and the gauge group $K$ collapses.  To prevent such a gauge symmetry breaking, $K$ must have a connected center.
\index{group ! center}%

(D)  \emph{Conjugacy class space with only central singularities.}  This is described in~\cite[p.~160]{froggatt}.

\medskip

These four conditions favour the Standard Model
\index{Standard Model (SM)}%
 group $\S(\U(2) \times \U(3))$
\index{group ! S(U(2) \times U(3))@$\S(\U(2) \times \U(3))$}%
 as surviving collapse in the gauge glass and surviving as a low energy symmetry group.  It is the one group whose Lie algebra
\index{Lie algebra}%
 has a non-trivial connected center.  It's one outer automorphism
\index{automorphism ! outer}%
 is complex conjugation.

\medskip

Regarding points (A) and (B) above, note that if $\alpha, \beta, \gamma \in \SO(8)$
\index{group ! SO(8)@$\SO(8)$}%
\index{group ! Spin(8)@$\Spin(8)$}%
 then $\Spin(8) \to \SO(8): \{(\alpha, \beta, \gamma), (- \alpha, - \beta, \gamma) \mapsto
\gamma\}$ is a $2$--$1$ surjection.  In fact one way of knowing that the Cartan triality
\index{Cartan triality}%
automorphisms (discussed at length in~\cite{tarzi1} and~\cite{tarzi2}) are outer is because they act nontrivially on the centre $\{(1,1,1),\ (1,-1,-1),\ (-1,1,-1),\ (-1,-1,1)\}$ of $\Spin(8)$.

\bigskip

The consistent string theories have either of the two gauge groups
\index{group ! gauge}%
 $\SO(32)$
\index{group ! SO(ua32)@$\SO(32)$}%
 or $\E_8 \times \E_8$,
\index{group ! E_8 \times E_8@$\E_8 \times \E_8$}%
 both of which have trivial centers.
\index{group ! center}%
  Group transformations act in pairs on the two ends of a string but elements of the center act as the identity.  Can the strings be considered as graph edges and if so would it then be meaningful to consider in a string context the physical meaning behind the action of groups supported by the random graphs?
\index{graph ! random ($\mathfrak{R}$)}%

\smallskip

The outer
\index{automorphism ! outer}%
 automorphic transposition of the two $\E_8$s in $\E_8 \times \E_8$
\index{group ! E_8 \times E_8@$\E_8 \times \E_8$}%
  may provide an alternative to the specific model applying the confusion mechanism
\index{confusion mechanism}%
 proposed as a global modification of the heterotic string model, wherein the two heterotic string $\E_8$s are permuted across a confusion hypersurface of codimension one representing a global topology
\index{topology}%
 change in space-time~\cite{benett4}, as a result of which the gauge group
\index{group ! gauge}%
 typically reduces to a single $\E_8$.
\index{group ! E_8@$\E_8$}%

\bigskip

\emph{Group Representation Theory}
\index{group ! representation theory}%

The standard model is based on the 12-dimensional algebra of $\SU(3) \times \SU(2) \times \U(1)$
 \index{group ! SU(3) \times SU(2) \times U(1)@$\SU(3) \times \SU(2) \times \U(1)$}%
and describes the forces between the gauge particles through a field theory.

The three generations of basic fermions are three instances of a specific collection of representations for this Lie algebra and thereby of the universal covering group $\SU(3) \times \SU(2) \times \mathbb{R}$,
 \index{group ! SU(3) \times SU(2) \times U(1)@$\SU(3) \times \SU(2) \times \mathbb{R}$}%
 from which it is obtained by factoring out a discrete
\index{group ! discrete}%
 normal
\index{group ! normal}%
 subgroup which is determined from a certain charge quantization rule~\cite{nielsenbrene}.

 The pattern of the fundamental fermion charges suggests the importance of a particular group having this algebra, namely $\S(\U(2) \times \U(3))$.
\index{group ! S(U(2) \times U(3))@$\S(\U(2) \times \U(3))$}%

Ever since the ground-breaking work of Wigner
\index{Wigner, E. P.}%
 group representation techniques have provided one of the most effective set of tools used in the classification of particle properties.  We mention just one such recent example, out of a vast literature going back over three-quarters of a century.

In looking to select a criterion as to why the SM gauge group
\index{group ! gauge}%
 turns out as it is, the one chosen in~\cite{nielsenhc} is the maximisation of a modified ratio of the orders of the quadratic Casimir for the adjoint representation over the quadratic Casimir of a representation chosen so as to make the quadratic Casimir the smallest possible non-trivial faithful representation.  The Lie gauge group
\index{group ! Lie}%
 would then have the smallest possible faithful representation in this sense of smallest, of the possible groups that gives the chiral Weyl fermions (leptons and quarks) and the Higgs boson.  In~\cite{nielsenhc}
\index{Nielsen, H. B.}%
 and in previous work with Bennett,
\index{Bennett, D. L.}%
Nielsen begins with a group that is the semi-direct product
\index{group ! semi-direct product}%
 of simple Lie groups
\index{group ! Lie}%
 by the additive group of real numbers; the number of abelian dimensions gives the dimension of the Lie algebra.
\index{Lie algebra}%
 Then he factors out of the center
\index{group ! center}%
 a normal discrete subgroup to get a group $G$.
\index{group ! discrete}%

The best known quadratic Casimir is the angular momentum operator $\overrightarrow{J}^2 = \overrightarrow{J}_x^2 + \overrightarrow{J}_y^2 + \overrightarrow{J}_z^2$ for $\SO(3)$.
\index{group ! SO(3)@$\SO(3)$}%
 The target quantity is
\[ \Big( \prod_i (\frac{C_A}{C_F})_{i}^{d_i} \star \prod_j (\frac{e_A^2}{e_F^2})_j \Big)^{\frac{1}{d_G}} \]
where the $i$ product runs over the normal simple groups with dimensions $d_i$, the $j$ product is over abelian factors, $d_G$ is the dimension of $G$, $e_F$ and $e_A$ are the ``charges'' for the $F$ and adjoint representations.  The word charges is in quotation marks because being abelian the $\mathbb{R}^+$ subgroups have no adjoint representations as the Lie algebra basis of a Lie group is only transformed trivially, so the second set of products in the above equation must be differently defined.  The ratio $\frac{C_A}{C_F}$ is related to the Dynkin index~\cite{difran}
\index{Dynkin index}%
 of a representation of a compact simple Lie algebra.
\index{Lie algebra}%

The target quantity has a value of at least 1, and is precisely 1 for both $\E_8$
\index{group ! E_8@$\E_8$}%
 and $\E_8 \times \E_8$
\index{group ! E_8 \times E_8@$\E_8 \times \E_8$}%
 gauge groups~\cite{nielsenhc}.
\index{group ! gauge}%
Of the groups tested, the target quantity attains its maximum for the SM group.
\index{group ! S(U(2) \times U(3))@$\S(\U(2) \times \U(3))$}%
The conclusion is that a principle of small representations should be enough to imply a significant part of the details of the SM.  

\bigskip

Despite the success of representation theory thus far the question can still be raised as to whether or not post-SM physics may not need new ideas and new approaches.

\bigskip

Our work has concentrated on the automorphism group $\Aut(\mathfrak{R^{t}})$
\index{group ! automorphism}%
 of the triality graph,
\index{graph ! triality ($\mathfrak{R^{t}}$)}%
 which being uncountably infinite in order, is less amenable to representation-theoretic techniques.  

However another criterion which is closer in spirit to our focus on the automorphism group of relational structures~\cite{tarzi2} than representation theory provides, and that distinguishes the group $\S(\U(2) \times \U(3))$
\index{group ! S(U(2) \times U(3))@$\S(\U(2) \times \U(3))$}%
  from all other connected compact non semisimple groups
\index{group ! semisimple}%
 with Lie algebra
\index{Lie algebra}%
 dimensionality up to 12 is that it has few \emph{generalised outer automorphisms},
\index{automorphism ! outer}%
in fact the least of any of the 192 connected non-semisimple groups with algebras of dimensionality 12 or less.  Brene and Nielsen
\index{Brene, N.}%
\index{Nielsen, H. B.}%
  proposed studying the orders $(n_i, n_j)$ of pairs of normal subgroups in the definition of generalized outer automorphisms in order to select the Standard Model group~\cite{nielsenbrene}, considering factoring out only discrete
\index{group ! discrete}%
 normal 
\index{group ! normal}%
subgroups of the center
\index{group ! center}%
 of a starting group.  Groups with simple algebras have finite discrete center and so a finite number of possible pairs, and those with continuous centers such as $\U(1)$
\index{group ! U(1)@$\U(1)$}%
 have an infinite number of level pairs.  Nonetheless some groups that do contain $\U(1)$s have few outer automorphisms because such groups are obtained from their covering group by factoring out a discrete normal subgroup with nontrivial projections in both the $\U(1)$ factor and other group factors, and so the original generalized $\U(1)$ symmetry is destroyed.

\bigskip

\emph{Groups with Triality and Triplication}
\index{group ! with triality}%

Despite the successes of the SM there are still around 20 parameters characterizing the couplings and particle masses which it does not explain.

The Family Replicated Gauge Group Model (FRGGM)
\index{Family Replicated Gauge Group Model (FRGGM)}%
 was proposed in~\cite{bennettniel} and~\cite{froggatt} as an extension of the SM.  Froggat and Nielsen
\index{Froggatt, C. D.}%
\index{Nielsen, H. B.}%
have presented~\cite{frogniel} a 5-parameter fit to the orders of magnitude of the quark-lepton masses and mixing angles in the FRGGM where the gauge group is $(\S(\U(2) \times \U(3)) \times \U(1)_{B - L})^3$
\index{group ! S(U(2) \times U(3))@$\S(\U(2) \times \U(3))$}%
 or otherwise written
\[ [\SU(3)_c]^3 \times [\SU(2)_L]^3 \times [\U(1)_Y]^3 \times [U(1)_{(B - L)}]^3.\]
\index{group ! SU(3)@$\SU(3)$}%
\index{group ! SU(2)@$\SU(2)$}%
\index{group ! U(1)@$\U(1)$}%
 In this extended SM model the
\index{Standard Model (SM)}%
 gauge group
\index{group ! gauge}%
 and a gauged $B - L$ (baryon number minus lepton number) is extended to one set of gauge fields per fermion family.  The 6 abelian gauge charges for the 3 families are the 3 weak hypercharges and the 3 $(B - L)$ charges, and these $\U(1)$
\index{group ! U(1)@$\U(1)$}%
 quantum charges generate the fermion mass hierarchy.  Higgs fields break down this family replicated gauge group to the SM group.  In~\cite{frogniel} a set of Higgs fields was constructed and VEVs were assigned to them, which fit all the quark-lepton masses and mixing angles, by taking the fundamental couplings to be of order unity and the fundamental masses to be of order the Planck mass; see also~\cite{das}.

\smallskip

The FRGGM approach is called \emph{anti-grand unified theory}.
 \index{anti-unification}%
 It assumes that space-time is discrete at very small distances and was developed as an alternative to SUSY GUTs.

There is a \emph{mass protection mechanism}, whereby fermions would have zero mass from gauge (charge) conservation, were it not for the Higgs mechanism.  This is thought to be due to the existence (of what are assumed to be gauge) quantum numbers lying outside the SM
\index{Standard Model (SM)}%
 which are assigned different values on the left-handed and right-handed Weyl components of the SM quarks and leptons.  It is hoped that an assignment of these chiral quantum numbers to the quarks and leptons can generate SM Yukawa coupling constants suppressed by the appropriate combinations of Higgs VEVs so as to be in agreement with experiment.  The FRGGM gauge quantum numbers are candidates for these mass-protecting chiral quantum numbers.

\bigskip

One of the key elements of the Random Dynamics program
\index{Random Dynamics}%
and the MPP is the speculation of a successive breaking of groups from a large over-group to for example $\S(\U(2) \times \U(3))^3$ or an extension thereof, and finally to the Standard Model
\index{Standard Model (SM)}%
 group $\S(\U(2) \times \U(3))$.
\index{group ! S(U(2) \times U(3))@$\S(\U(2) \times \U(3))$}%

This final step is reminiscent of the phenomenon of \emph{Groups with Triality},
\index{group ! with triality}%
discussed in~\cite{tarzi1}, which is a generalization of Cartan
\index{Cartan triality}%
 $\Sym(3)$
\index{group ! Sym(3)@$\Sym(3)$}%
outer automorphism
\index{automorphism ! outer}%
 triality of the group $\SO(8)$.
\index{group ! SO(8)@$\SO(8)$}%
 and the three representations of its double cover spin group $\Spin(8)$.
\index{group ! Spin(8)@$\Spin(8)$}%

\smallskip

We conjectured the possibility of a unified triality whereby the embedding of three two-colour random graphs $\mathfrak{R}$
\index{graph ! random ($\mathfrak{R}$)}%
 in one triality graph $\mathfrak{R^{t}}$, 
\index{graph ! triality ($\mathfrak{R^{t}}$)}%
 three copies of the spin group $\Spin(7)$
\index{group ! Spin(7)@$\Spin(7)$}%
 in one spin group $\Spin(8)$,
\index{group ! Spin(8)@$\Spin(8)$}%
 three $\mathbb{L}_{E_8}$
\index{lattice ! mathbb{L}_{E_8}@$\mathbb{L}_{E_8}$}%
 lattices in a Leech lattice $\mathbb{L}_{L}$,
\index{lattice ! Leech}%
 and finally Lie Algebras with Triality,
\index{Lie algebra with triality}%
 are all related phenomena.  

We refer to~\cite{tarzi1} and~\cite{tarzi2} for discussions of these topics.

\bigskip

In~\cite{tarzi2} we motivated an argument for how the 3-family horizontal structure might arise from the triality graph,
\index{graph ! triality ($\mathfrak{R^{t}}$)}%
 with the colours $\mathfrak{r, b, g}$ identifying the generations -- \emph{generational quantum numbers}.
\index{quantum number}%

\smallskip

A different model inspired by Random Dynamics was given by D. L. Bennett, H. B. Nielsen and I. Picek
\index{Bennett, D. L.}%
\index{Nielsen, H. B.}%
\index{Picek, I.}%
 in~\cite{bennettniel}.

\chapter{Summary}

Random Dynamics
\index{Random Dynamics}%
 takes as its starting point the idea that the most advantageous assumption that we can make about the fundamental laws is that they are random.  That a fundamental theory is so beyond our reach that to all intents and purposes we can take its laws to be random.  Furthermore that the known laws of relativity and quantum mechanics emerge from this fundamental theory.  Despite its successes the project has thus far this project has been a niche endeavour.   By identifying a candidate as the sought after random structure we hope to increase interest in this line of enquiry.

\smallskip

The countably infinite random graph
\index{graph ! random ($\mathfrak{R}$)}%
 $\mathfrak{R}$ is the unique (up to isomorphism) connected structure that encodes the connections of every possible configuration of sets of vertices.  Multicoloured versions of the graph are natural extensions.  The graph $\mathfrak{R}$ can in fact be reconstructed from its automorphism group $\Aut(\mathfrak{R})$ (as can the multi-coloured versions~\cite{rubin})
\index{graph ! random ! $m$-coloured}%
\index{group ! Aut(\mathfrak{R})@$\Aut(\mathfrak{R})$}%
\index{group ! automorphism}%
 so that in the right context the group can be thought of as a `prior' construct.

\smallskip

Symmetries of the physical laws that seem to govern the wokings of nature are not in general symmetries of the world because the latter are subject to \emph{initial conditions}, which are not invariant under symmetries of nature's laws~\cite{wigner}.  It would appear that many of the well-known symmetries arise naturally out of the SM.  The $\mathcal{C}\mathcal{P}\mathcal{T}$ Theorem
\index{CPT@$\mathcal{CPT}$ Theorem}%
 can be derived for any quantum field theory from a few assumptions~\cite[Chapter~5]{froggatt}.

Poincar\'e
\index{group ! Poincar\'e}%
 and gauge invariance
\index{gauge ! invariance}%
 (and perhaps translation invariance) must be added separately and so these are regarded as the fundamental symmetries~\cite[pp.~42-43]{froggatt}.  Nonetheless there have been attempts to derive even these, classified according to three different methods~\cite[Chapter~6]{froggatt}, (i) formal dynamical derivations that reveal the symmetry which then takes on a physical interpretation in some vacuum
\index{vacuum}%
 phase due to quantum fluctuations, (ii) from the renormalisation group in the limit of larger distances or lower energies, and (iii) as a consequence of string theory properties.

The automorphism group $\Aut(\mathfrak{R^{t}})$ 
\index{group ! Aut(\mathfrak{R^{t}})@$\Aut(\mathfrak{R^{t}})$}%
of the 3-coloured (triality) random graph
\index{graph ! triality ($\mathfrak{R^{t}}$)}%
 is connected to the modular group
\index{group ! modular}%
 $\PSL(2, \mathbb{Z})$
\index{group ! PSL(2, \mathbb{Z})@$\PSL(2, \mathbb{Z})$}%
 through the construction of $\mathfrak{R^{t}}$~\cite[Theorem~6.6]{tarzi1} as a homogeneous Cayley graph
\index{graph ! Cayley}%
for an index $3$ subgroup of the modular group
\index{group ! modular}%
\[ \PSL(2, \mathbb{Z}) = \langle \sigma, \rho : \sigma^{2} = \rho^{3} = 1 \rangle.\]
\index{group ! PSL(2, \mathbb{Z})@$\PSL(2, \mathbb{Z})$}%

Modular groups of fractional linear transformations acting on the upper half-plane in $\mathbb{C}$ take the form
\[ z \mapsto \frac{az + b}{cz + d},\quad a, b, c, d \in \mathbb{Z}, \mathbb{R}. \]

By taking $\mathbb{C}$ instead of $\mathbb{Z}$ as the normed division algebra for the modular group
\index{group ! modular}%
 it is a short step by complexification to its Lorentz over-group
\index{group ! Lorentz}%
 $\PSL(2, \mathbb{C})$
\index{group ! PSL(2, \mathbb{C})@$\PSL(2, \mathbb{C})$}%
 and we arrive at Lorentz invariance
\index{Lorentz invariance}%
 which is thus, as required by the Random Dynamics program, derived from a random structure without having to be assumed a priori.

$$
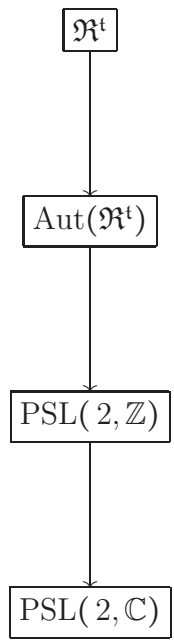
\xymatrix{ 
*+[F]{\mathfrak{R^{t}}}  \ar[dd]\\
\\
*+[F]{\Aut(\mathfrak{R^{t}})} \ar[dd]\\
\\
*+[F]{\PSL(2, \mathbb{Z})} \ar[dd]\\ 
\\
*+[F]{\PSL(2, \mathbb{C})}
}$$
\index{group ! Aut(\mathfrak{R^{t}})@$\Aut(\mathfrak{R^{t}})$}%
\index{group ! PSL(2, \mathbb{Z})@$\PSL(2, \mathbb{Z})$}%
\index{group ! PSL(2, \mathbb{C})@$\PSL(2, \mathbb{C})$}%

\bigskip

An intermediate group in the complexification is the group of conformal automorphisms
\index{group ! automorphism}%
 of the upper half-plane $\PSL(2, \mathbb{R})$
\index{group ! PSL(2, \mathbb{R})@$\PSL(2, \mathbb{R})$}%
 of which the modular group $\PSL(2, \mathbb{Z})$
\index{group ! modular}%
\index{group ! PSL(2, \mathbb{Z})@$\PSL(2, \mathbb{Z})$}%
is a discrete subgroup.  
\index{group ! discrete}%

The upper half-plane can be thought of as a hyperbolic plane.  In~\cite{feingold} and~\cite{kleinschmidt} Feingold, Kleinschmidt, Nicolai and Palmkvist
\index{Feingold, A.}%
\index{Kleinschmidt, A.}%
\index{Nicolai, H.}%
\index{Palmkvist, J.}%
 study isomorphisms between hyperbolic Weyl groups
\index{group ! hyperbolic Weyl}%
 and modular groups
\index{group ! modular}%
 over integer domains in normed division algebras:
\[ \PSL(2, \mathbb{R}) \cong SO_0(2, 1) \]
\[ \PSL(2, \mathbb{C}) \cong SO_0(3, 1) \]
\[ \PSL(2, \mathbb{H}) \cong SO_0(5, 1) \]
\[ \PSL(2, \mathbb{O}) \cong SO_0(9, 1) \]
\index{group ! PSL(2, \mathbb{H})@$\PSL(2, \mathbb{H})$}%
\index{group ! PSL(2, \mathbb{O})@$\PSL(2, \mathbb{O})$}%
 where for example, $\PSL(2, \mathbb{R})$
\index{group ! PSL(2, \mathbb{R})@$\PSL(2, \mathbb{R})$}%
 is the identity component of the Lorentz group
\index{group ! Lorentz}%
 of 3-dimensional spacetime, acting on the hyperboloid
\[ \{ (t, x, y) \in \mathbb{R} : t^2 - x^2 - y^2 = 1, t > 0 \} \]
 this being one representation of the hyperbolic plane.

The isomorphism between the Lorentz group $\SO(3, 1)$
\index{group ! SO(3, 1)@$\SO(3, 1)$}%
 and the complexified rotation group $\SO(3, \mathbb{C})$
\index{group ! SO(3, \mathbb{C})@$\SO(3, \mathbb{C})$}%
 is the basis of the formalism that lies behind various self-dual formulations of general relativity~\cite{krasnov}, such as that of
\index{Pleba\'nski, J. F.}%
 Pleba\'nski~\cite{plebanski} which uses self-dual two-forms instead of the dynamics of a metric, and of which the Hamiltonian formulation of general relativity due to Ashtekar~\cite{ashtekar}
\index{Ashtekar, A.}%
 is the phase space version as proved by Jacobson and Smolin~\cite{jacobson}.
\index{Jacobson, T.}%
\index{Smolin, L.}%
 A model has been constructed by Froggatt, Das, Laperashvili, Nielsen and Tureanu~\cite{frogz}
\index{Das, C. R.}%
\index{Laperashvili, L. V.}%
\index{Froggatt, C. D.}%
\index{Nielsen, H. B.}%
\index{Tureanu, A.}%
 which uses this formulation as well as the Multiple Point Principle
\index{Multiple Point Principle (MPP)}%
 to unify gravity with some weak gauge fields.

The \emph{integer octonions} $\textbf{O}$ form a discrete subring of the octonions $\mathbb{O}$ and is modulo a rescaling, just the $\E_8$ root lattice which we denote $\mathbb{L}_{E_8}$,
\index{lattice ! mathbb{L}_{E_8}@$\mathbb{L}_{E_8}$}%
 this being the smallest even unimodular lattice embedded within an 8-dimensional Euclidean space.
\index{Euclidean space}%
  The discrete
\index{group ! discrete}%
 subgroup $\PSL(2, \textbf{O})$ of $\PSL(2, \mathbb{O}) \cong SO_0(9, 1)$
\index{group ! SO(9, 1)@$\SO(9, 1)$}%
 acts on the 9-dimensional hyperbolic space $\{ x \in \mathbb{R}^{10} : x^2_0 - x^2_1 - \ldots x^2_9 = 1, x_0 > 0 \}$, where $SO_0(9, 1)$ is the identity component of the Lorentz group
\index{group ! Lorentz}%
 of 10-dimensional Minkowski spacetime.

The above four identifications must account for the noncommutativity of both quaternions $\mathbb{H}$ and  octonions $\mathbb{O}$, and nonassociativity
\index{nonassociativity}%
 of $\mathbb{O}$.  Various `integer' subrings of the division algebras are described together with their relationships to hyperbolic Coxeter groups~\cite{feingold}.
\index{group ! Coxeter}%

An example of nonassociativity
\index{nonassociativity}%
 in physics is the special relativistic velocity composition law that relates the velocities of moving objects in different reference frames.  The velocities add as if they are hyperbolic tangent ($\tanh$) functions because the Lorentz transformation can be thought of as the application of a hyperbolic rotation through a hyperbolic angle (which is velocity divided by $c$).  This is used in the gyrovector space description of Thomas precession, where in Abraham Ungar's
\index{Ungar, A.}%
 theory gyrovector addition is based on a type of Bol loop called a \emph{gyrogroup}.

\medskip

 We see again that the rotation groups
\index{group ! orthogonal (rotation)}%
 and modular groups
\index{group ! modular}%
 are fundamental, just as we emphasized in~\cite{tarzi2}.  As indeed are lattices; Feingold et al studied a type of Weyl group
\index{group ! Weyl}%
 of hyperbolic Kac-Moody algebras
\index{Kac-Moody algebra}%
 which are intimately linked with the four normed division algebras $\mathbb{R}, \mathbb{C}, \mathbb{H}, \mathbb{O}$, making crucial use of integral lattices as well as the integer subrings of the division algebras and associated discrete
\index{group ! discrete}%
\index{group ! matrix}%
 matrix groups.

We constructed~\cite{tarzi1} $\mathfrak{R^{t}}$
\index{graph ! triality ($\mathfrak{R^{t}}$)}%
 as a homogeneous Cayley graph
\index{graph ! Cayley}%
for the complex Leech Lattice
\index{lattice ! complex Leech}%
as a special case of a construction of random Cayley $m$-coloured graphs
\index{graph ! random ! $m$-coloured}%
 from lattices $\mathbb{L}$
\index{lattice}%
 in $\mathbb{R}^d$, for $m, d \ge 2$, identifying lattice vectors with graph vertices and vector pairs with edges such that the graphs are invariant under the lattice automorphism group $\Aut(\mathbb{L})$.  This particular link may have an application to global (that is, non-space-time dependent) symmetry properties of gauge glass
\index{gauge ! glass}%
 theoretic models.

\medskip

In~\cite{chkareuli}, Chkareuli, Froggatt and Nielsen
\index{Chkareuli, J. L.}%
\index{Froggatt, C. D.}%
\index{Nielsen, H. B.}%
 argue that Lorentz non-invariant
\index{Lorentz invariance}%
 effects caused by the spontaneous breakdown of Lorentz symmetry (SBLS) are physically unobservable, and that application of this principle to the most general relativistically invariant Lagrangian, with arbitrary couplings for all the fields, gives rise to both abelian and non-abelian massless vector gauge fields.  More specifically to the appearance of a continuous symmetry in terms of conserved Noether currents and to the massless vector fields gauging this symmetry.  In other words, gauge invariant abelian and non-abelian theories can be derived from the requirement of the physical non-observability of the SBLS rather than by using the Yang-Mills gauge principle.  Matrices of coupling constants form Lie algebra
\index{Lie algebra}%
 representations corresponding to this symmetry, and in the non-abelian case the coupling constants for the vector field self- interaction terms are the structure constants for this algebra.  Therefore the vector gauge fields become a source of the symmetries, rather than local symmetries being a source of the gauge fields as in the usual formulation.  Imposing the condition of unobservability of SBLS restricts the values of the coupling constants and mass parameters in the Lagrangian density.  The issue of which mechanisms could induce the SBLS is taken up in~\cite{chkareuli1}, where it is argued that the non-observability of the SBLS caused by the vacuum expectation values (VEV)
\index{vacuum expectation value (VEV)}%
 of scalar gauge fields could provide the origin of all observed internal symmetries.  
The presence of scalar fields can possibly induce non-zero classical fields as the VEVs of the original vector fields.  The conclusion is that a gauge symmetry phase is created in order to avoid a breakdown of Lorentz invariance.

Even global (non-gauge) symmetries are not required in the original Lagrangian because the SBLS  induces them automatically as accidental symmetries accompanying the generated gauge theory.  

In an older reference, assuming only the classical electrodynamics of a charged particle interacting with an electromagnetic plane wave field, Kupersztych
\index{Kupersztych, J.}%
 derives~\cite{kupersztych} a Lorentz-transformation operator that describes the motion of the spin of a particle having a magnetic moment precisely that of a Dirac spin particle, and is simultaneously a gauge transformation (which thus leaves invariant a plane wave field).  This connects gauge invariance, relativistic invariance and electron spin which has no classical analogue.  However, the derivation assumes the Lorentz force law which is obeyed by any charged particle, including the pion which has no spin.

\bigskip

Yang-Mills theory is a gauge theory based on the $\SU(n)$
\index{group ! SU(n)@$\SU(n)$}%
 group but can more generally be based on any compact, semisimple
\index{group ! semisimple}%
 Lie group.
\index{group ! Lie}%
\index{group ! compact}%
\index{group ! semisimple}%
For example, electroweak and strong force theories are respectively based on $\SU(3)$ and $\SU(2) \times \U(1)$.
\index{group ! SU(3)@$\SU(3)$}%
\index{group ! SU(2) \times U(1)@$\SU(2) \times \U(1)$}%
 A purely formal derivation of gauge symmetry
\index{gauge ! symmetry}%
 appears in~\cite[p.~110]{froggatt} and as a consequence of Lorentz invariance in the model of~\cite{chkareuli}.  If Lorentz invariance
\index{Lorentz invariance}%
 can be derived or arrived at from a random discrete mathematical structure, whether through our purported route or any other, is there an effective discrete content to group-theoretic  gauge invariance?
\index{gauge ! invariance}%

Consider the well-known subgroup relationship between orthogonal and unitary
\index{group ! unitary}%
 groups $O(n) < \U(n) < O(2n)$, and in particular that $\SO(n) < \SU(n)$.  Being compact
\index{group ! compact}%
 the orthogonal
\index{group ! orthogonal (rotation)}%
 group has discrete
\index{group ! discrete}%
 subgroups that are equivalent to finite subgroups, amongst which is the Coxeter group
\index{group ! Coxeter}%
 $A_{n-1} \cong \Sym(n)$
\index{group ! symmetric ($\Sym$)}%
 of permutation matrices (which are themselves orthogonal matrices).

Note also that to get from $O(n)$ to its subgroup $O(n-1) < O(n)$ simply consider rotations that stabilize an axis; this simple fact allows us to go down the orthogonal group series.
\index{group ! orthogonal (rotation)}%
 Further that $O(n)/\SO(n) \cong O(1)$ with the projection map picking matrices with $\pm 1$ determinants.  Orthogonal matrices with determinant -1 cannot include the identity, so form a coset rather than a subgroup.  The projection map splits so $O(n)$ is the semi-direct product
\index{group ! semi-direct product}%
 $\SO(n) \sd O(1)$.  Two references for the classical groups are~\cite{porteous} and~\cite{taylor}.

The matrices representing $\SO(n)$ have determinant $+1$ describing \emph{proper rotations} rather than rotations combined with reflections.  But the determinant of a permutation matrix equals the signature of the corresponding permutation.  So there are natural maps from $O(n) \to \Sym(n)$, and $\SO(n) \to \Alt(n)$.
\index{group ! alternating ($\Alt$)}%

So there are maps $U(3) \to O(3) \to \Sym(3)$, $U(2) \to O(2) \to \Sym(2) \cong C_2$ and $U(1) \to O(1) \to \Sym(1) = id$.  

\smallskip

The following identifications are known:\\
$\PSL(2,2) = \PGL(2, 2) \cong \Sym(3)$, $\PGL(2, 3) \cong \Sym(4)$, $\PSL(2,3) \cong \Alt(4)$, $\PSL(2,4) \cong \PSL(2,5) \cong \Alt(5)$.
\index{group ! alternating ($\Alt$)}%
\index{group ! symmetric ($\Sym$)}%
Is there a series of mappings with an enlightening physical interpretation, from $\PSL(2, \mathbb{Z})$
\index{group ! modular}%
\index{group ! PSL(2, \mathbb{Z})@$\PSL(2, \mathbb{Z})$}%
 to the Standard Model
\index{Standard Model (SM)}%
 group $\S(\U(2) \times \U(3))$?
\index{group ! S(U(2) \times U(3))@$\S(\U(2) \times \U(3))$}%

\medskip

There is an interest in these types of questions that make use of the important discrete ingredients in continuous groups.  One such is the Froggatt-Nielsen mechanism
\index{Froggatt-Nielsen mechanism}%
 for explaining the hierarchies of fermion masses and mixings in terms of a $\U(1)$
\index{group ! U(1)@$\U(1)$}%
 symmetry~\cite{froggattnielm}. 

The fermion masses and mixings are given by Yukawa couplings of the form $Y^{ij} \overline{\psi}_i \psi_j H$, for two interacting fermionic fields, a Higgs boson and a Yukawa matrix.  For their proper description we could either fine-tune correlations amongst the entries of the Yukawa matrix $Y$ or we could assume the basic (bare) coupling constants to be random (but at small distances are all of the same order of magnitude) and that the entries of $Y$ itself contain a hierarchical pattern.  But what would be the origin of this randomness?  The mechanism proposed introduces so-called \emph{flavon}
\index{flavon}%
 $\U(1)$ charged fields $\phi$, which are neutral under the SM gauge group
\index{group ! gauge}%
 but that like the Higgs, develop a vacuum expectation value (VEV) to generate mass,
\index{vacuum expectation value (VEV)}%
 via the introduction of an expansion parameter $\epsilon = \langle \phi \rangle / \Lambda$ where
$\Lambda$ denotes a high energy mass scale.

The three generations of left-handed and right-handed quark fields (up-type quarks are not distinguished from down-type quarks) carry different charges under $\U(1)_{FN}$ such that the usual Yukawa terms have positive integer charges.   The hierarchy in the Yukawa matrix entries arises from powers of the flavon field compensating for these non-zero charges of the above trilinear Yukawa terms.  Imposition of this mechanism reduces the number of free parameters if the new $\U(1)$
\index{group ! U(1)@$\U(1)$}%
 family symmetry is spontaneously broken down to a residual discrete $\mathbb{Z}_N$ symmetry.  The article~\cite{king} further take up the topic of flavons.

\bigskip

The uniqueness of the multicoloured random graphs
\index{graph ! random ! $m$-coloured}%
 on a given set of colours in the infinite-vertex limit starting from arbitrary initial configurations would seem to suggest that the physical interpretation of any result derived therefrom would be independent of initial conditions. 

One physical interpretation that we have in mind is this: that the growth of a graph from an arbitrary initial configuration to the infinite random graph
\index{graph ! random ($\mathfrak{R}$)}%
 is mirrored by taking the limit of a large number of particles.  There are many instances of limits of large numbers in physics, such as the hydrodynamical scaling law for an incompressible fluid being derived when the number of fluid molecules becomes very large~\cite[p.~8]{froggatt}.

\medskip

The intimate relationship between the modular
\index{group ! modular}%
 and 3-braid groups, for example that the 3-braid group
\index{group ! braid}%
 $B_3$ is the universal central extension of the modular group~\cite{tarzi2}, offers a direction for further study.

\medskip

A key test for any theory that extends the SM model will be in its ability to better explain the SM
\index{Standard Model (SM)}%
 coupling constants and mass parameters.  It is possible that quantum field theory will only be able to do this through the introduction of a new ingredient such as a supplementary symmetry.  The hope has been that application of the MPP
\index{Multiple Point Principle (MPP)}%
 gives information about these SM parameters.  It is also possible that such additional restricting principles are not a priori clearly or directly related to the dynamical entities and only deeper work would reveal the connections.  In~\cite{tarzi2} we have proposed candidates and have tried to motivate them.   It could be argued that the restrictions should, for reasons of simplicity, apply to the zero-particle states, that is the vacua.  

\smallskip

Given that Random Dynamics
\index{Random Dynamics}%
 is based upon an entirely different philosophy to the the more mainstream approaches of unification, we end with a philosophical point regarding choice of the best theory from alternatives, though within philosophy itself there is no consensus on an accepted theory of this subject.

Science operates through observation, analysis, reason, deduction and sometimes prediction.  It functions according to a precision, which is why mathematics is often required for its proper explanation and description.  Scientific theories are never complete; in physics this has led to the rise of so-called ``effective theories''.  As Wells has pointed out~\cite{wellsjd}
\index{Wells, J. D.}%
to all intents and purposes all theories are effective theories, choosing Thagard's criteria~\cite{thagard}
\index{Thagard, P.}%
 for \emph{best explanation}:

1.  \emph{Consilience:} How many facts the theory explains, and furthermore unifies and systematizes.

2.  \emph{Simplicity:} Having the least number of auxiliary hypotheses and additions.

3.  \emph{Analogy:} A theory explaining the shared characteristics between competing theories by admitting a new characteristic with which to explain the shared ones.

Observational and mathematical consistency with known empirical facts are sine qua non and take precedence over other factors.  Conjecture is probably the most effective source of theories and knowledge grows by rearrangement combination, alteration and subsuming of previous principles.  Criticism and empiricism allows choice of theories from alternatives.  Good explanations have robustness~\cite{deutschbeg}. 

To this end the entire Random Dynamics
\index{Random Dynamics}%
 program may be one of the most economical possibilities, if it could be developed to anywhere near the ambitious intentions of its founders, beginning with specific random structures.  We have made a starting suggestion as to what such structures might be.

\printindex

\end{document}